\documentclass[11pt]{article}
\usepackage{moriond,epsfig}

\def\be{\begin{equation}}
\def\ee{\end{equation}}
\def\bea{\begin{eqnarray}}
\def\eea{\end{eqnarray}}

\newcommand{\MET}{$\not\!\!E_T$}
\def\pt{$p_T$}
\def\Aboson{$A$}
\def\Hboson{$H$}
\def\hboson{$h$}
\def\bbbar{$b\bar{b}$}

\def\tanb{$\tan{\beta}$}

\def\btag{$b$}
\def\Missing#1#2{{\mbox{$#1\kern-0.57em\raise0.19ex\hbox{/}_{#2}$}}}

\def\vMissing#1#2{\ifmmode
            \vec{#1}\kern-0.57em\raise.19ex\hbox{/}_{#2}
         \else
            {{\mbox{$\vec{#1}\kern-0.57em\raise.19ex\hbox{/}_{#2}$}}}
         \fi}
\def\lsim{\mathrel{\rlap{\lower4pt\hbox{\hskip1pt$\sim$}}
    \raise1pt\hbox{$<$}}}        
\def\gsim{\mathrel{\rlap{\lower4pt\hbox{\hskip1pt$\sim$}}
    \raise1pt\hbox{$>$}}}

\def\pt{\mbox{$p_{T}$}}

\def\D0{D\O }

\newcommand{\rar}{\rightarrow}
\def\simge{\mathrel{\rlap{\raise 0.53ex \hbox{$>$}}%
{\lower 0.53ex \hbox{$\sim$}}}}
\def\simle{\mathrel{\rlap{\raise 0.53ex \hbox{$<$}}%
{\lower 0.53ex \hbox{$\sim$}}}}
\begin{document}
\vspace*{4cm}
\title{Searches for Higgs Boson(s) at the Upgraded  Tevatron}

\author{Gregorio Bernardi \\ (for the CDF and D\O\ Collaborations)}

\address{LPNHE - Paris VII and VII, and Fermi National Accelerator Laboratory, \\
{\it{gregorio@in2p3.fr}}}

\maketitle\abstracts{
We summarize the status of Higgs boson searches at the upgraded Fermilab Tevatron 
performed by the D\O\ and CDF collaborations.
We report on three categories of searches, namely the search for the Standard Model 
Higgs boson ($p \bar{p} \rar H, \ WH$ or $ZH$, with $H \rar WW^*$ and/or $H \rar b \bar{b}$), the 
search for the minimal supersymmetric Higgs boson using $p\bar{p} \rightarrow hb\bar{b} \rightarrow 
b\bar{b}b\bar{b}$ and $p\bar{p} \rightarrow hX \rightarrow \tau\tau X$, and the search for
 doubly charged Higgs boson.
}

\section{The Standard Model Higgs Boson}

\begin{figure}[t]
\psfig{figure=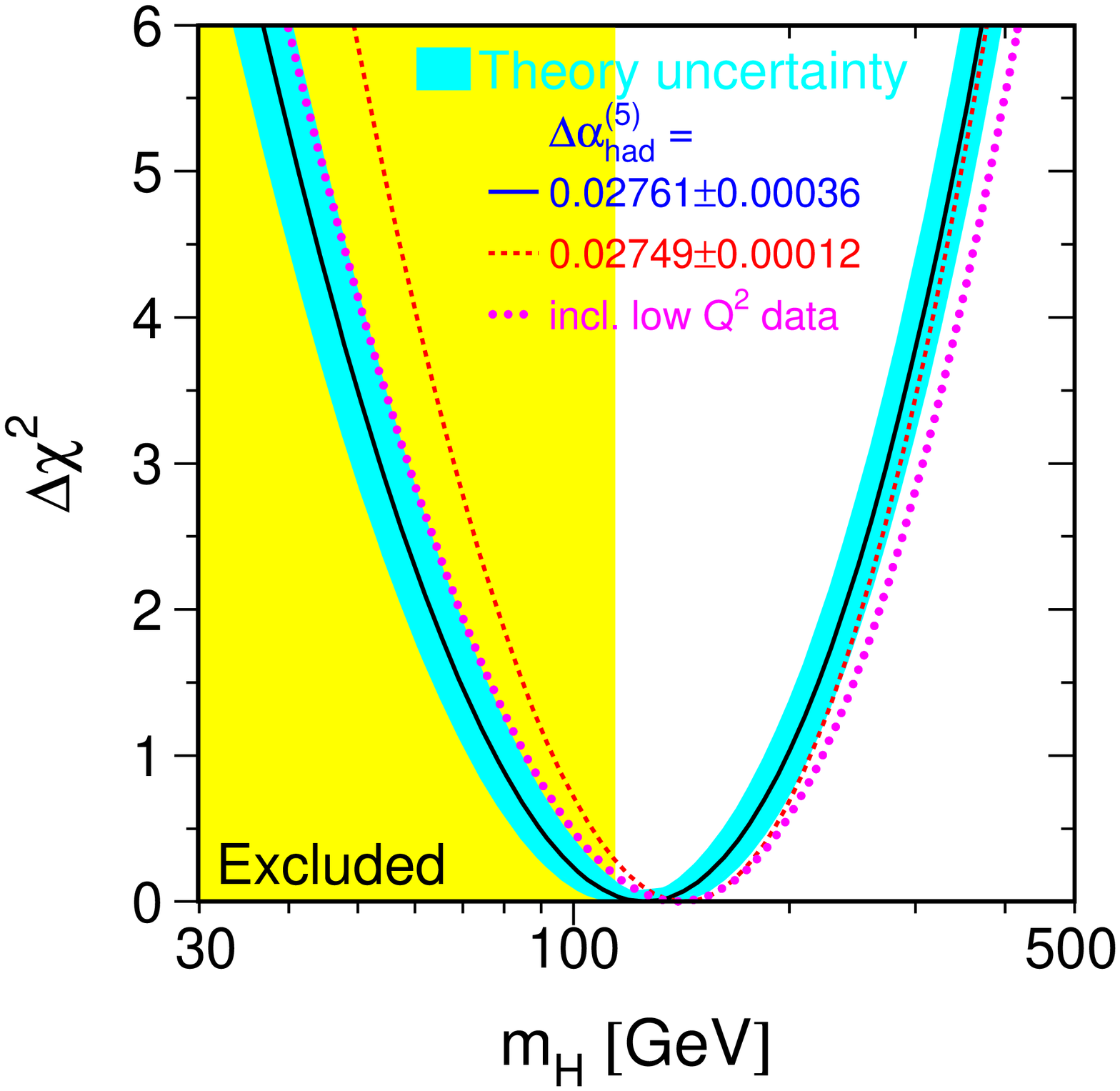,height=2.2in,width=2.0in}
\psfig{figure=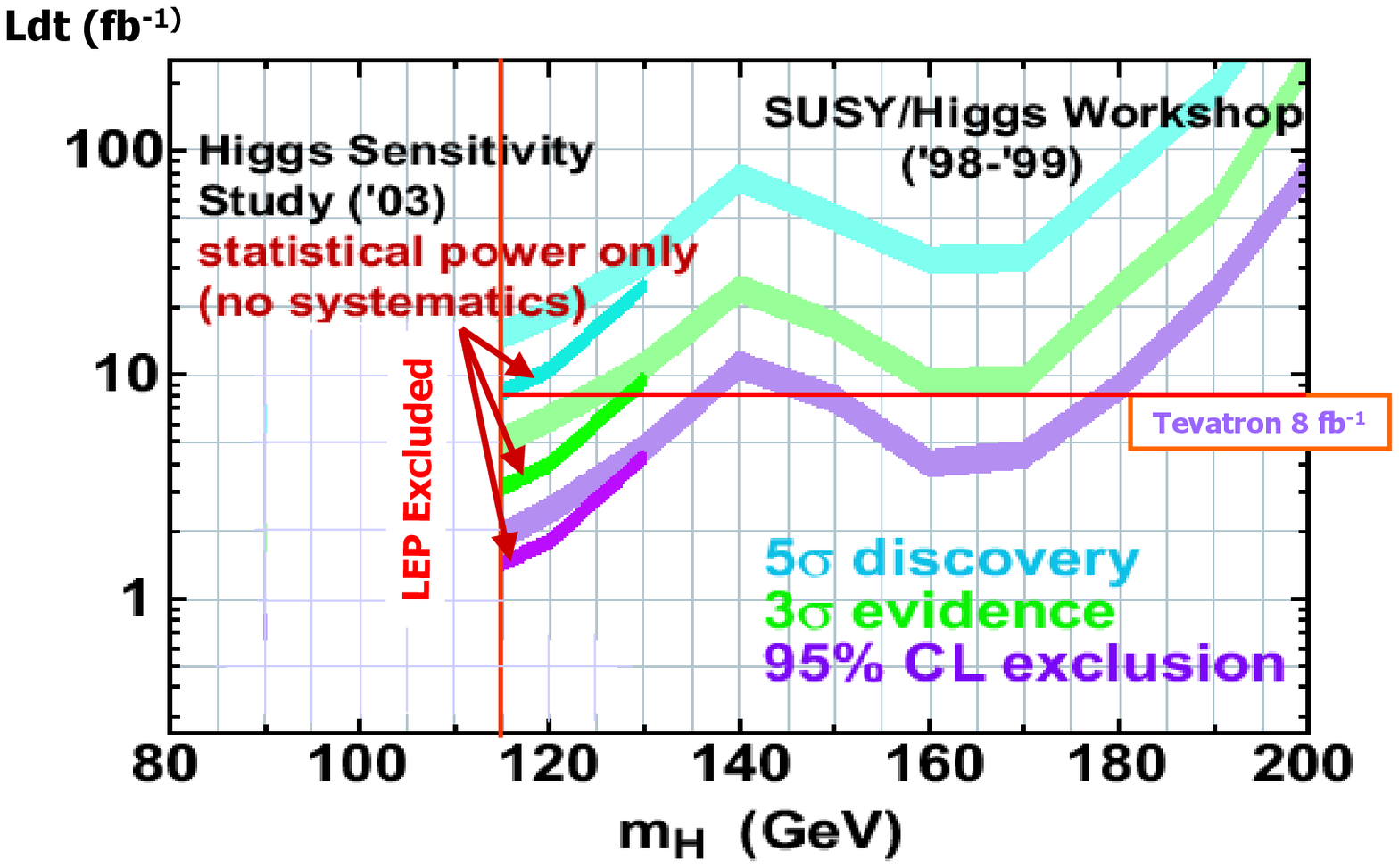,height=2.0in,width=2.5in}
\psfig{figure=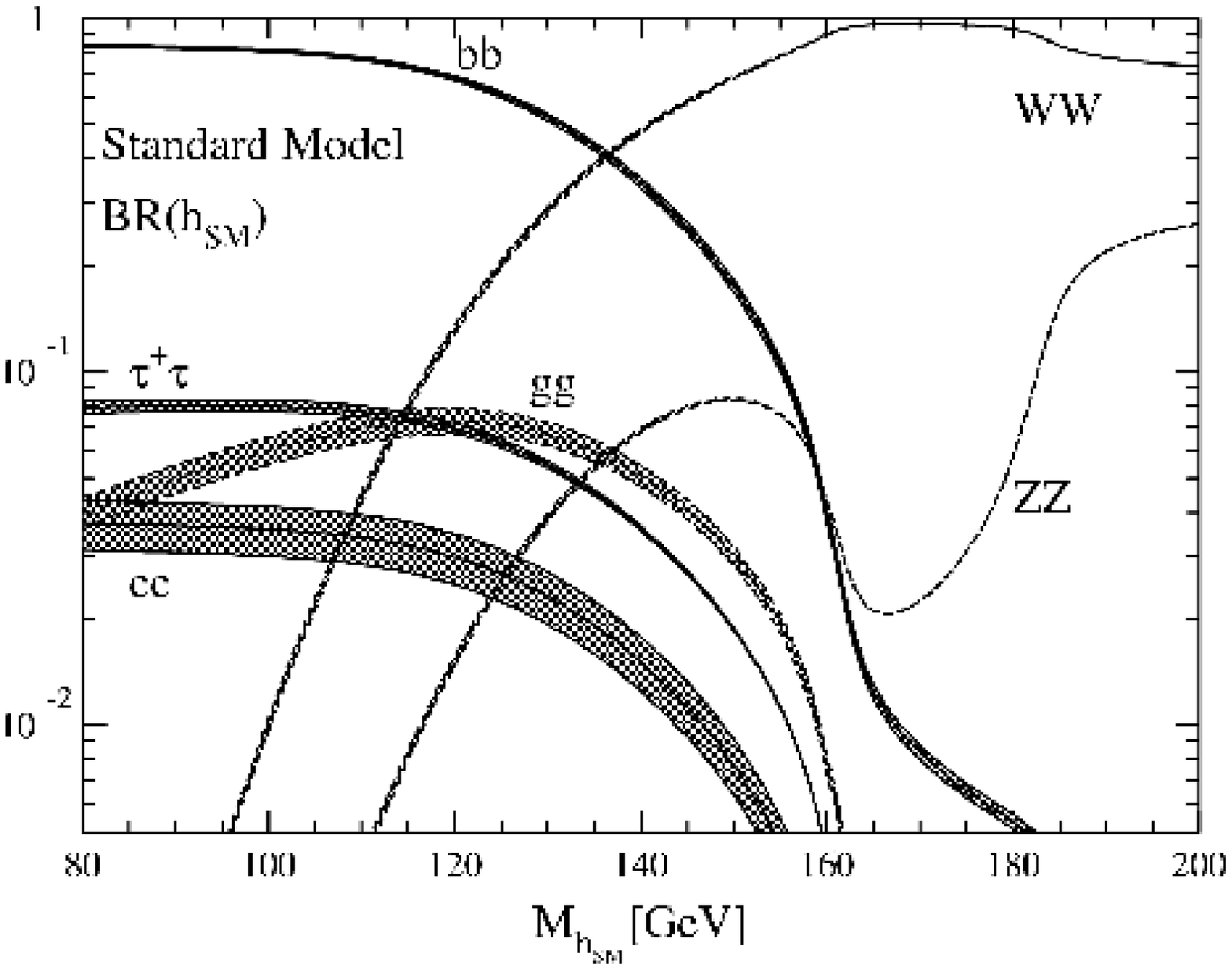,height=2.0in,width=1.6in}
   \caption{
a) Constraints on the Higgs mass from a fit to Electroweak measurements; 
b) Expected sensitivity to the Higgs boson at the Tevatron; 
c) Branching ratios ($BR$) of the Higgs boson as a function of $m_H$, in the SM.
\label{fig:ewwgSMH}}
\end{figure}

The Higgs boson is the only scalar elementary particle expected in the 
standard model (SM). Its discovery would be a major success for the SM and 
would provide new experimental insights into the electroweak symmetry breaking
mechanism. 
Direct measurements at LEP have excluded a SM Higgs boson with a mass 
$m_H < 114.4$ GeV at 95\% C.L. but
constraints from precision measurements  nevertheless favor a 
Higgs boson sufficiently light to be accessible at the Fermilab Tevatron 
Collider. 
The current preferred mass value, as deduced from a fit to 
electroweak measurements by the LEP, SLD, 
CDF, and D\O\ experiments~\cite{ewwg} shown in Fig.~\ref{fig:ewwgSMH}a,
is  $126^{+73}_{-48}$ GeV.
At the Tevatron, indirect searches involve 
precision measurements of the top quark and the $W$ mass, while 
direct searches require high luminosity samples for discovery or exclusion in 
the $115-185$ GeV mass range, as shown in Fig.~\ref{fig:ewwgSMH}b.
Although the expected luminosity necessary for its discovery at the 
Tevatron is higher than 
obtained thus far, the special role of the Higgs boson in the SM justifies
extensive searches for a Higgs-like particle independent of expected 
sensitivity. The chances for discovery  are better in
supersymmetric (SUSY) extensions of the SM~\cite{susy}, since the
SUSY Higgs production cross-section is generally predicted to be larger than that of the SM.


\section{Searches for SM Higgs Production at the Upgraded Tevatron}

At the Tevatron 
$p\overline{p}$ collider ($\sqrt{s}$=1.96~TeV), the 
two dominant mechanisms for Higgs production are 
gluon fusion, and associated production with a $W$ or a $Z$:
$gg \rightarrow H$, $q\overline{q} \rightarrow W/Z + H$. 
Although the $gg$ process has the largest cross section, $\sim$~1~pb at
$m_{H}$=115~GeV, 
it is the most sensitive production mode only at relatively high mass Higgs boson,
($m_H \gsim$130~GeV) where it has a significant branching ratio in $WW^{*}$, as shown
in Fig.~\ref{fig:ewwgSMH}c. At lower masses, the dominant decay $H \rar b\bar{b}$ is 
swamped by multijet background, so only the search for a Higgs produced in assocation
with a vector boson has enough sensitivity.
The $WH$ and $ZH$ channels, would give a clear signal 
with lepton(s), neutrino(s), and 2 $b$-jets: 
$q\overline{q} \rightarrow W/Z+H \rightarrow l\nu /ll/\nu\nu +b\overline{b}$.

Two main studies  have been performed on 
the sensitivity of the Tevatron experiments to SM Higgs physics. 
The first one, in 98-99~\cite{susy-higgs},
explored the whole mass range available with some approximation of the detector response,
while the second one, in '03, was restricted to the low mass region~\cite{higgsens} and was
using a more realistic simulation, since first data of Tevatron Run II had become available.
The most recent study essentially confirmed the findings of the original study, and both
results
are summarized in Fig.~\ref{fig:ewwgSMH}b, after combination of all channels of both
experiments. The amount of integrated luminosity needed for a Higgs
discovery at $m_H =$ 115~GeV is approximately 8~fb$^{-1}$. A 3 $\sigma$ evidence might
be found with $\gsim 3-4 $fb$^{-1}$, while most of the Higgs mass region below
$\sim$185~GeV
could be excluded at 95\% CL with $\sim 8$~fb$^{-1}$. 
\begin{figure}[htbp]
\psfig{figure=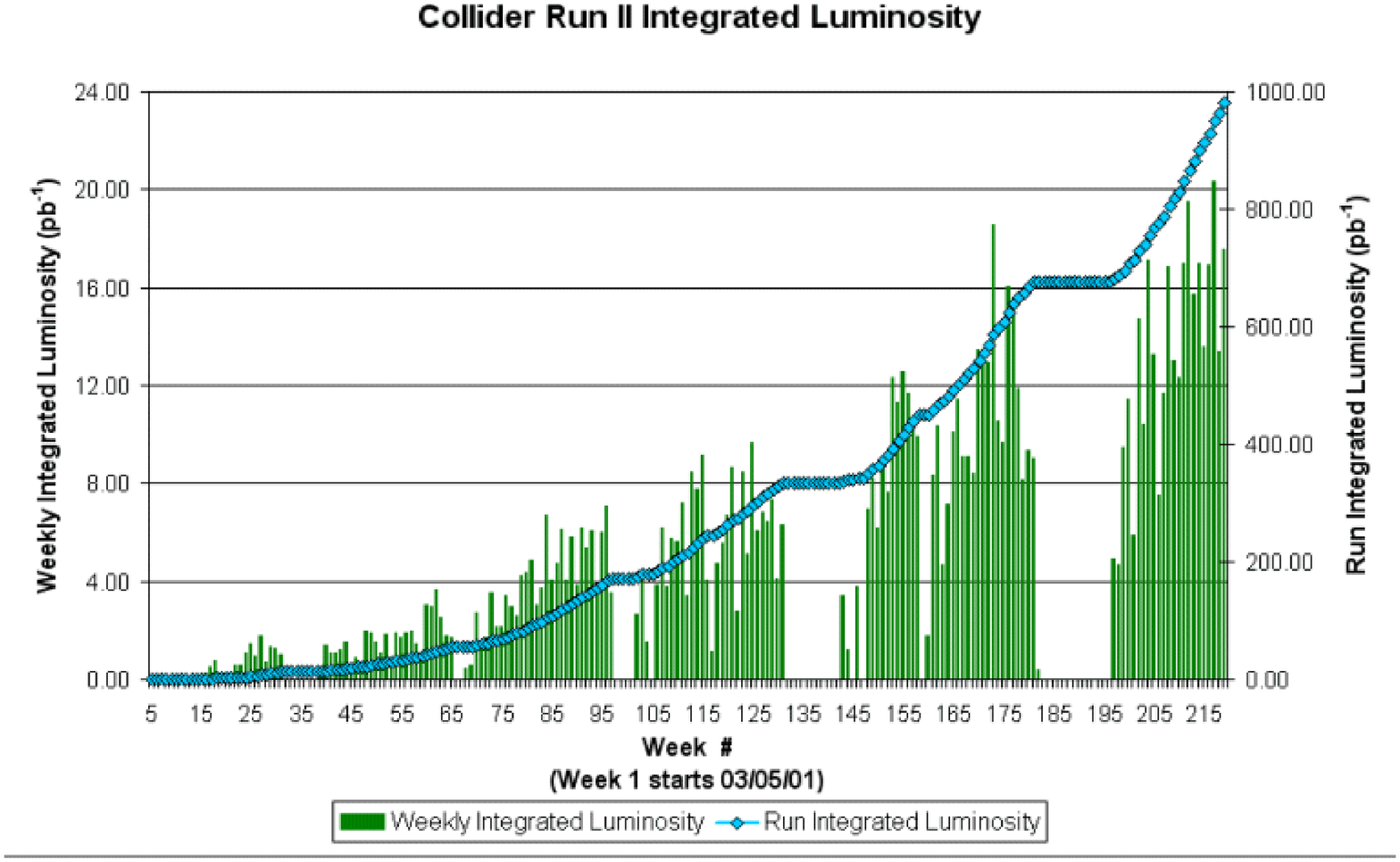,width=2.1in,height=2.2in
,bbllx=0,bblly=10,bburx=764,bbury=524,clip=}
\psfig{figure=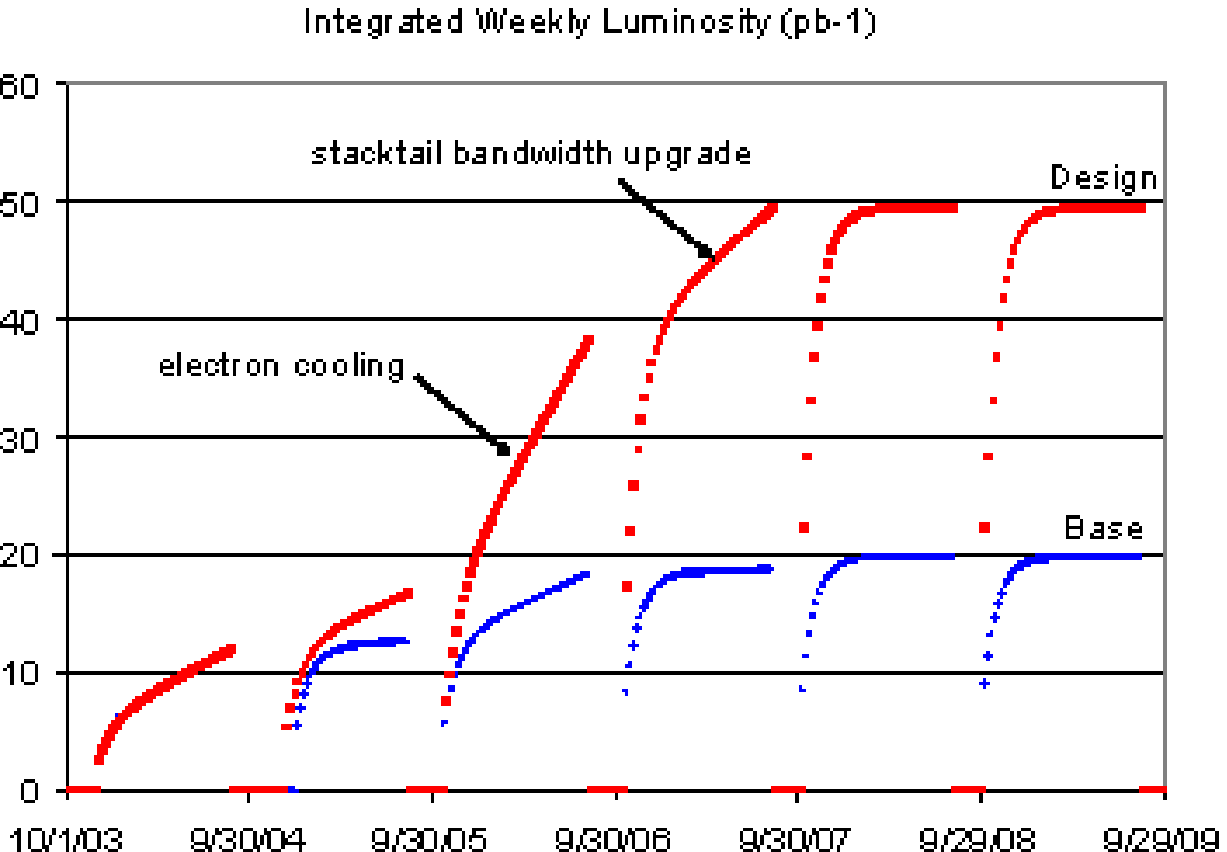,width=2.2in,height=2.2in
,bbllx=-40,bblly=-40,bburx=390,bbury=290,clip=}
\psfig{figure=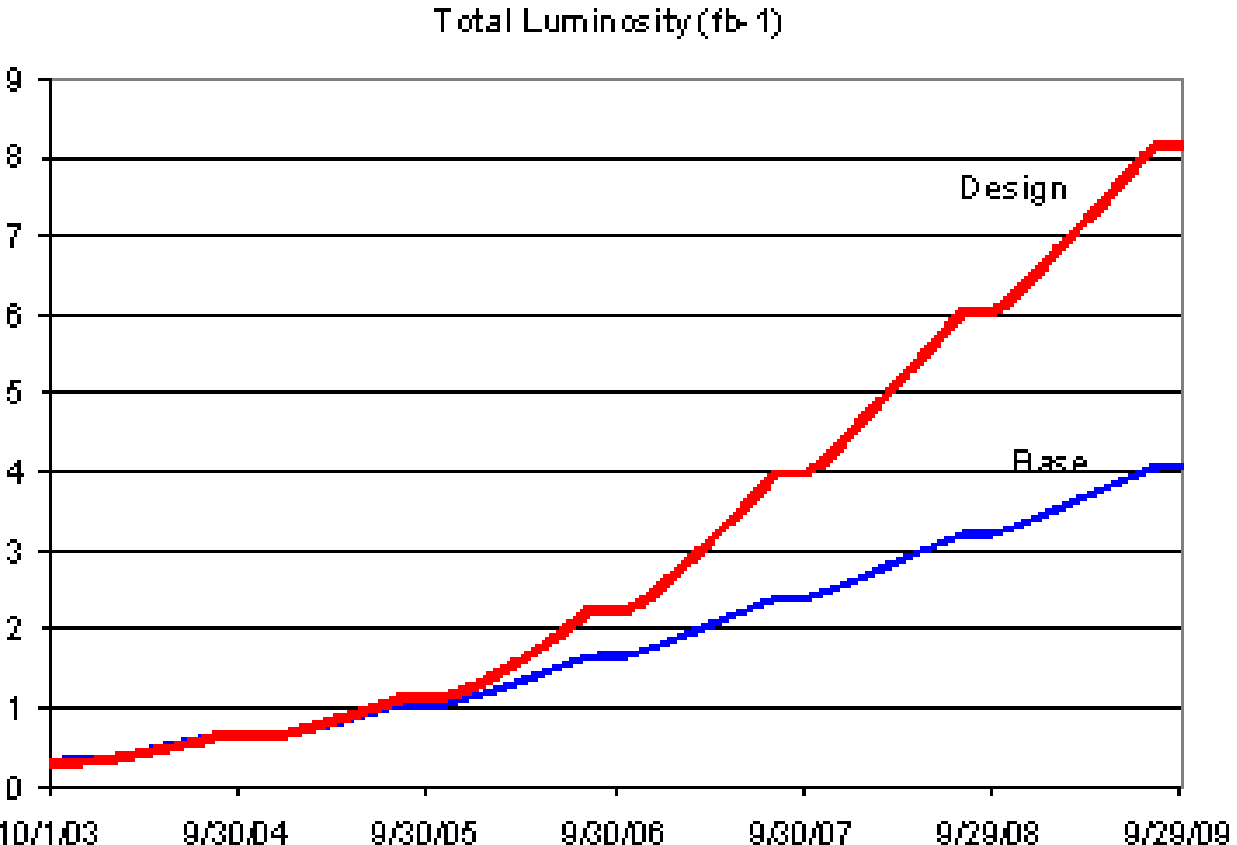,width=2.2in,height=2.2in
,bbllx=-40,bblly=-40,bburx=390,bbury=280,clip=}
  \caption{
   Current and expected performance of the Tevatron: a) integrated luminosity
and weekly delivered luminosity since the beginning of Run II; 
b) Expected weekly delivered luminosity;  c)
Expected integrated luminosity.
    \label{fig:tev} }
\end{figure}

These searches are crucially dependent on the performance of the Tevatron.
After a "slow" start, the machine is performing very well since the end of 2003, 
with larger than
designed delivered luminosities. Currently about 1 fb$^{-1}$ has been delivered as shown
in Fig.~\ref{fig:tev}a, 
with weekly integrated luminosity which have reached over 
20 pb$^{-1}$. 
Since this figure is the terminal value assumed in the minimal (base)
luminosity expectation (cf Fig.~\ref{fig:tev}b), the current performance ensure that,
barring accident, a minimal integrated luminosity of 4 fb$^{-1}$ will be achieved by
the end of 2009, as shown in Fig.~\ref{fig:tev}c. If the accelerator keeps following
the designed luminosity evolution, a luminosity of $8.2$ fb$^{-1}$ will be achieved,
rendering the potential for a Higgs discovery significant at the Tevatron. 

\subsection{Search for  Higgs Production in 
Association with a $W$ boson, in the $W b \bar{b}$ Final State}

\begin{figure}[htbp]
\psfig{figure=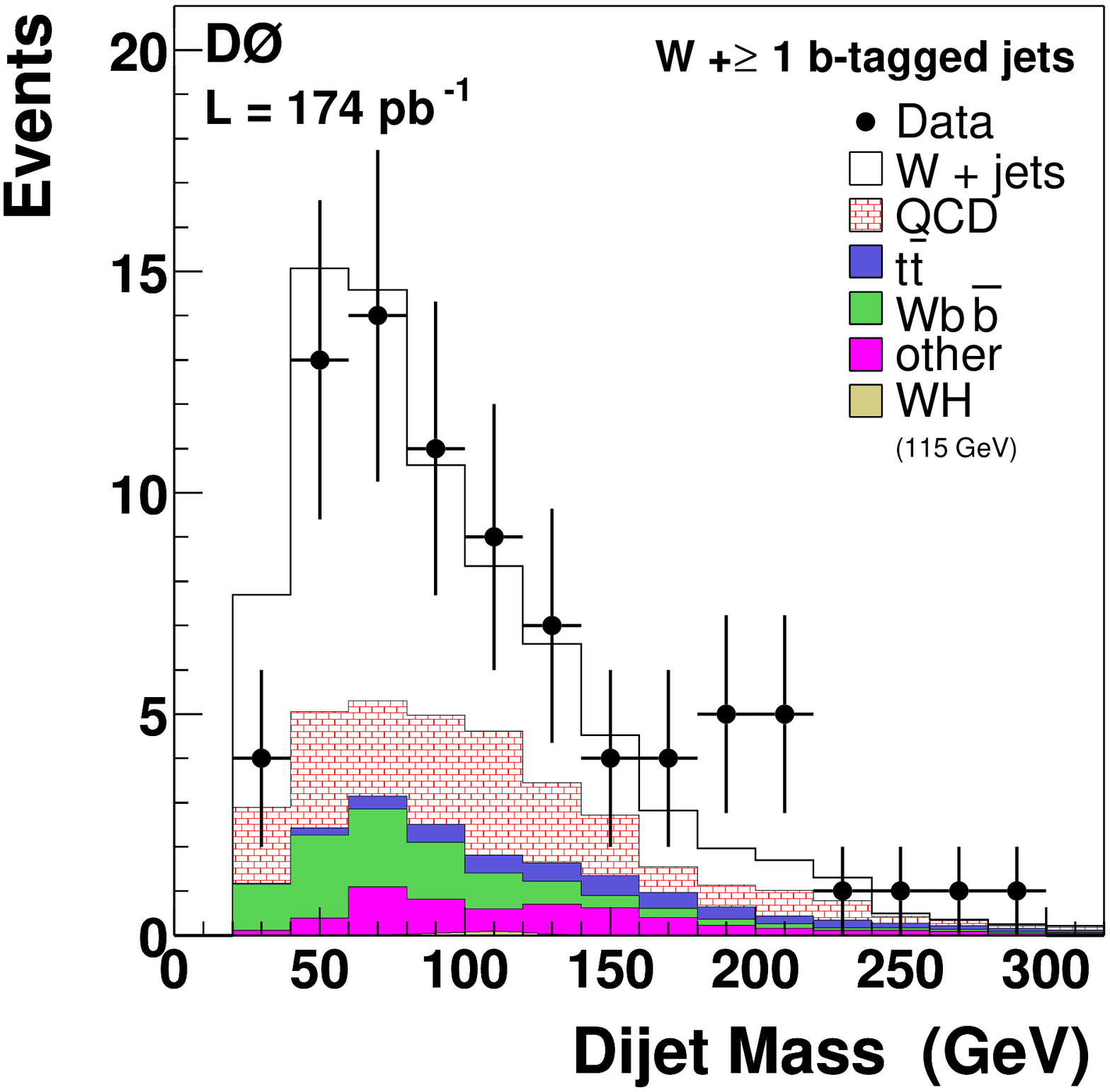,height=2.0in}
\psfig{figure=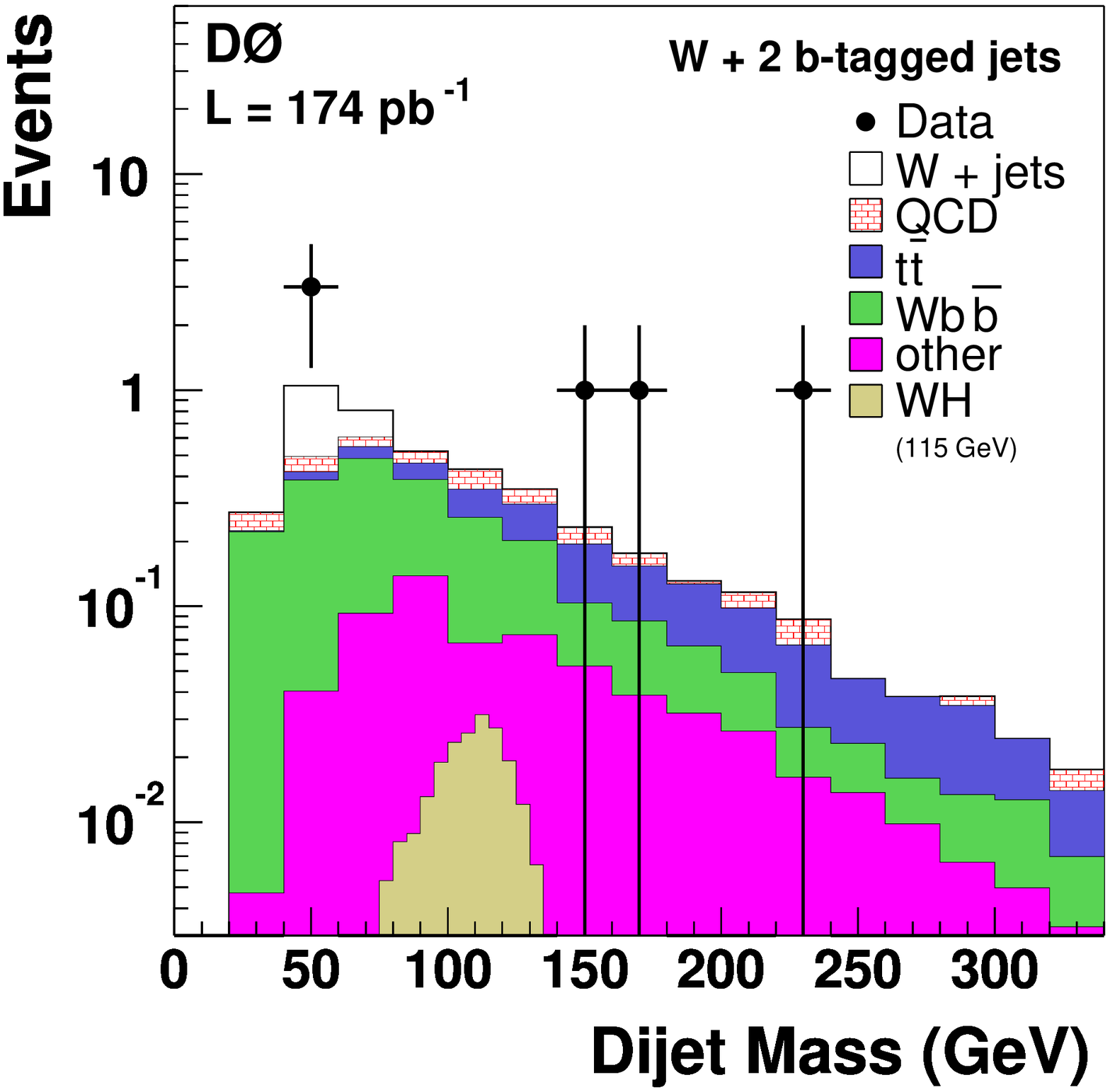,height=2.0in}
\psfig{figure=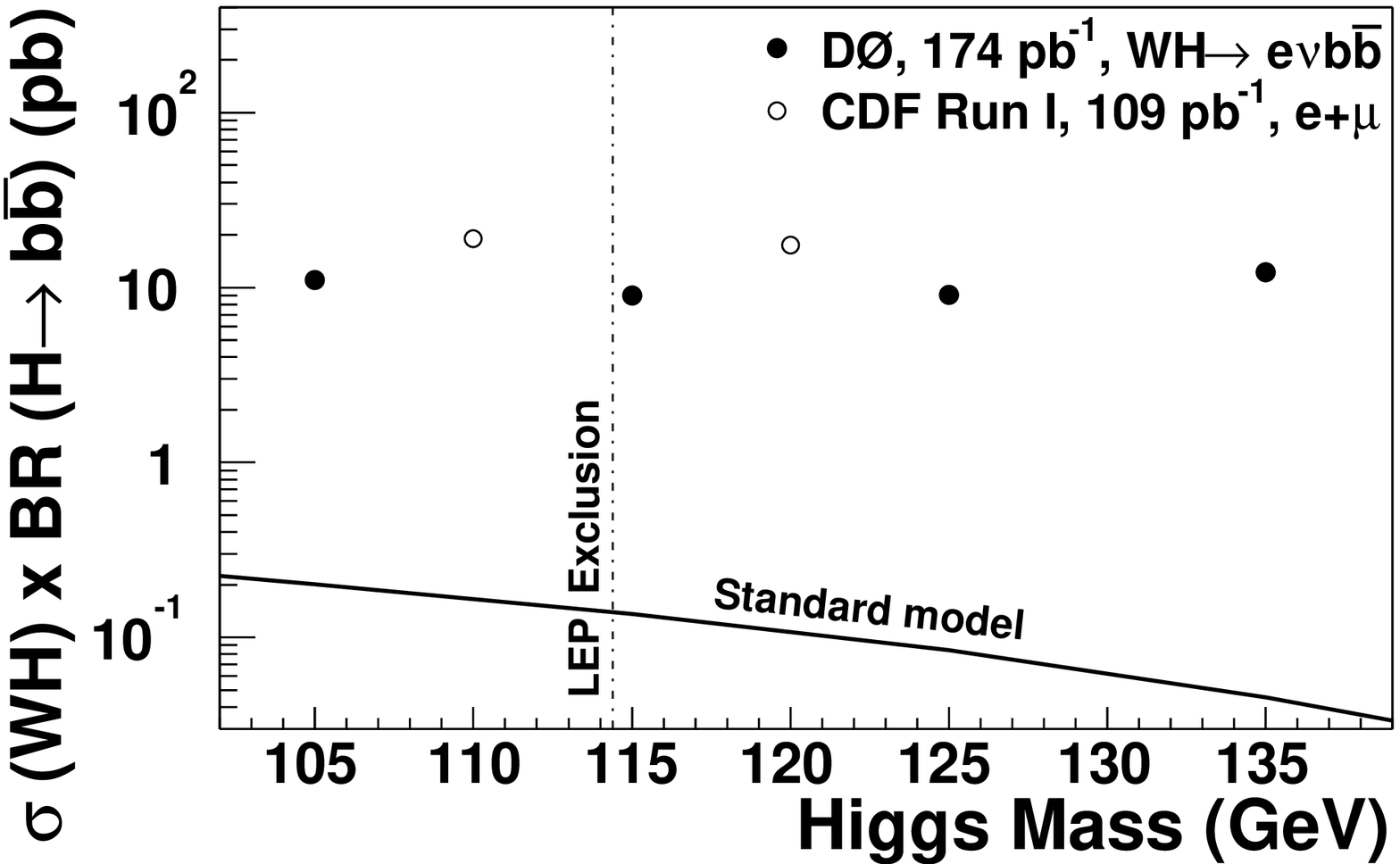,height=2.0in,width=2.2in}
  \caption{
WH results at D\O.
a) Distribution of the dijet invariant mass in the $W+ $ 2 jets sample after requiring at least one
$b$-tagged jet b)
Same distribution, after requiring that the 2 jets are $b$-tagged. The expectation obtained
from a 115 GeV Higgs is also shown. c)  Limit on
$WH~\times~BR(H \rar b \bar{b})$
cross section, as a function of $m_H$. 
     \label{fig:wh-d0} }
\end{figure}

CDF and D\O\ have studied the  $Wb\bar{b}$ final state, in which $WH$ production
could be observed.
The D\O\ analysis, based on
174 pb$^{-1}$ of data,
requires two $b$-tagged jets with 
$p_T >$ 20 GeV and $\eta < 2.5$, large transverse missing energy
(\MET $> 25$ GeV) and an isolated electron with 
$p_T >20$ GeV. The measured dijet mass distribution is well described by the
expectation, both before (not shown) and after single $b$-tagging (shown in 
Fig.~\ref{fig:wh-d0}a).  The number of observed events is reduced from
 76 (2 jets including at least 1 $b$-tagged jet),
to 6 events when requesting 2 $b$-tagged jets (Fig.~\ref{fig:wh-d0}b). 
The number of expected  events being 4.4, a 95\% C.L. upper cross section limit  
of 6.6  pb is set on  $Wb\bar{b}$ production for $p_T^b >20$ GeV and an $\eta-\varphi$
separation between $b$-jets $>$~0.75. By restricting the selection to 
a $\pm$25 GeV window around the searched Higgs masses, 
D\O\ has established a 95\% C.L. limit on $WH$ 
production of 9$-$12.2 pb for $m_H$ between 105 and 135 GeV~\cite{prl-wh-wbb}, 
as shown in Fig.~\ref{fig:wh-d0}c. 

Although this limit is still significantly above the expected SM cross section,
the difference in sensitivity compared to 
the '03 prospective study, shown in Fig.~\ref{fig:ewwgSMH}b, is ``only''
a factor 2.4, 
when taking into account the current luminosity and this specific channel only.
 This factor is understood for the major part of it, since i) 
the current analysis is temporarily restricted to ``central'' electron ($\eta <1.1$), 
while the final coverage, up to 2.5, was used in ~Ref.~\cite{higgsens}; 
ii) a ``tight'' $b$-tagging algorithm is currently applied,
while new developments show that a significant gain can be achieved in $b$-tagging
efficiency by loosening it; 
iii) several other efficiencies (trigger, lepton-id) can be improved. 
We expect thus that 
after implementing the analysis improvements mentioned above,
the sensitivity prospects will be met.

\begin{figure}[htbp]
\psfig{figure=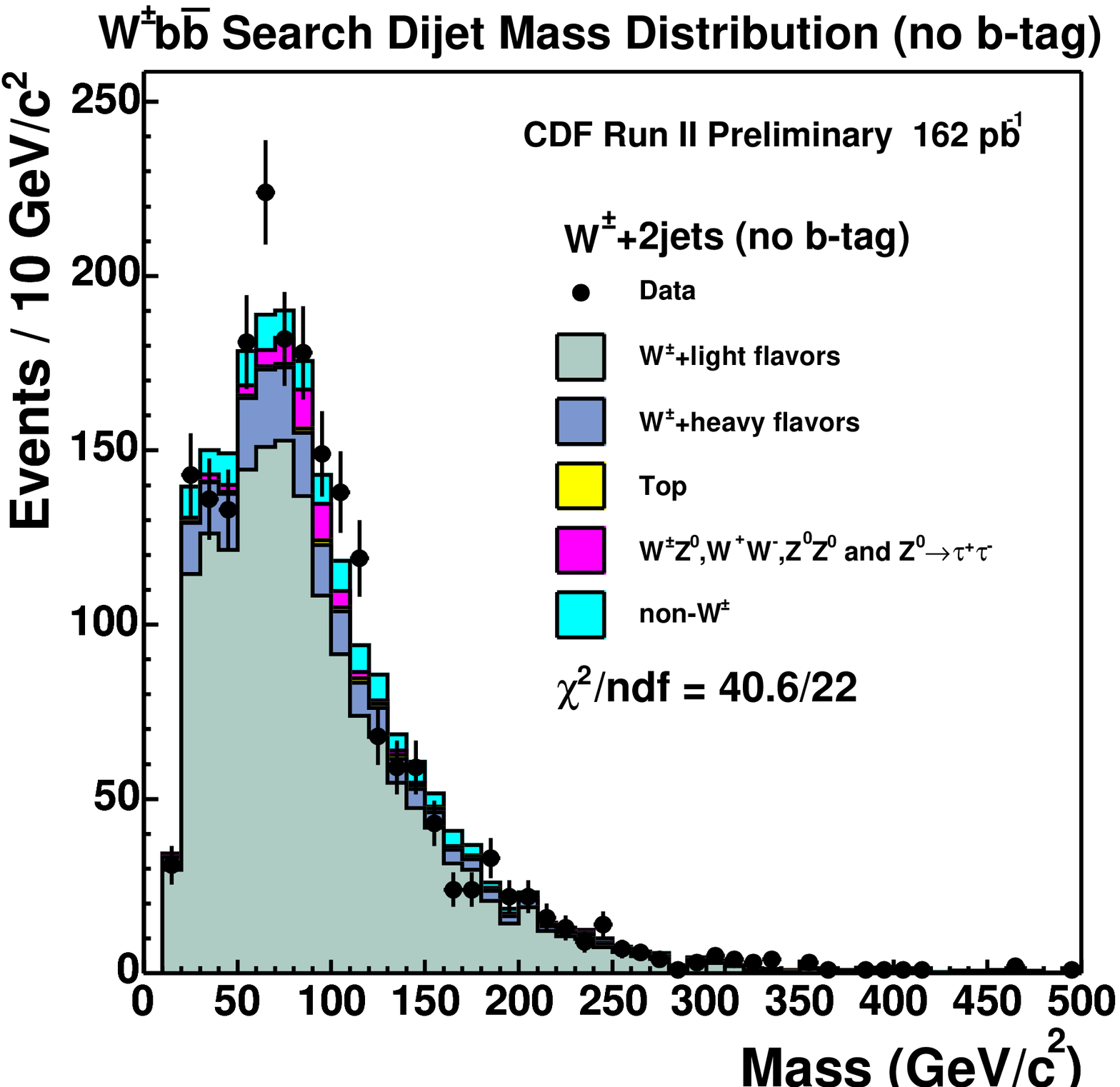,height=2.0in}
\psfig{figure=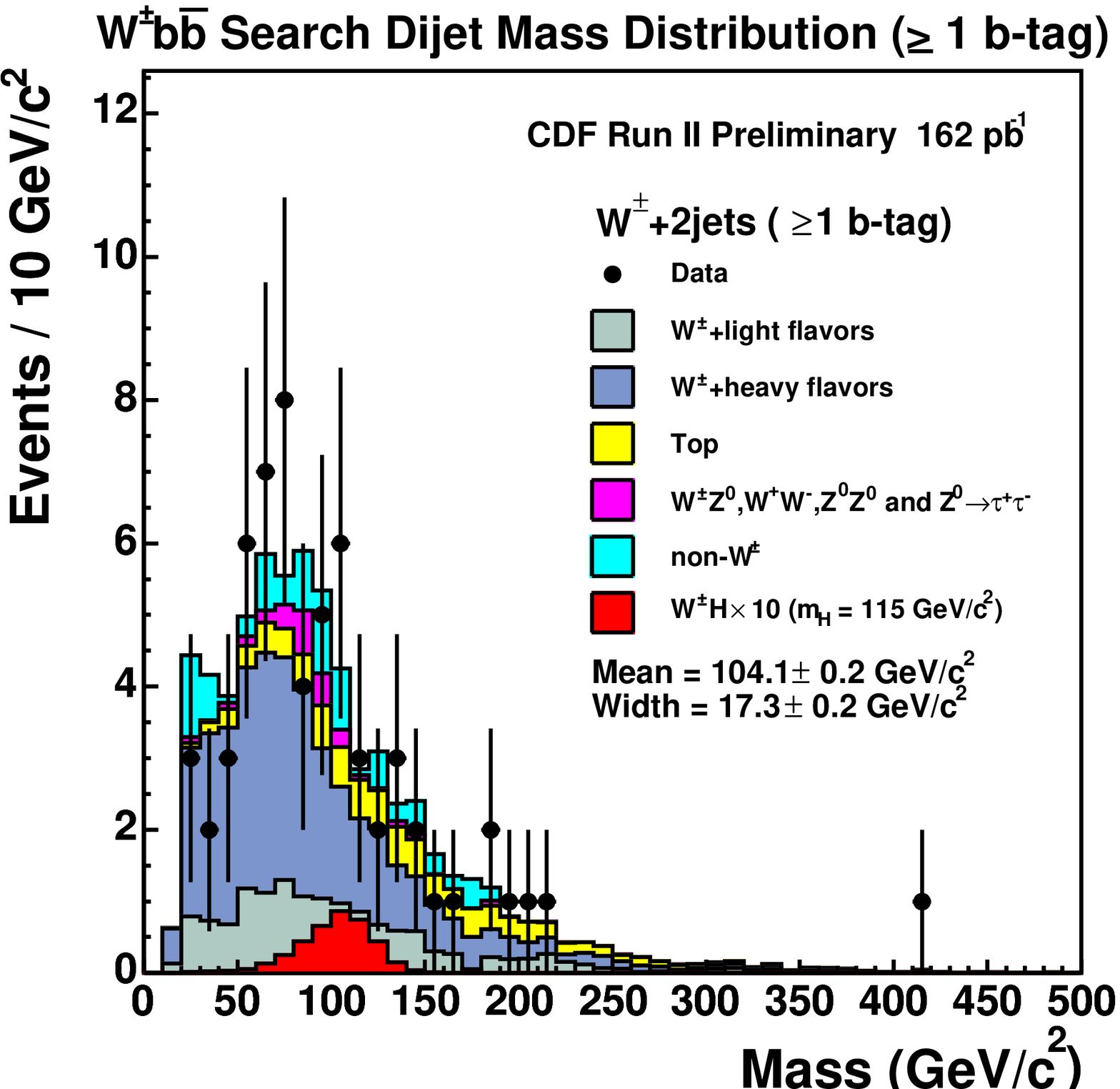,height=2.0in}
\psfig{figure=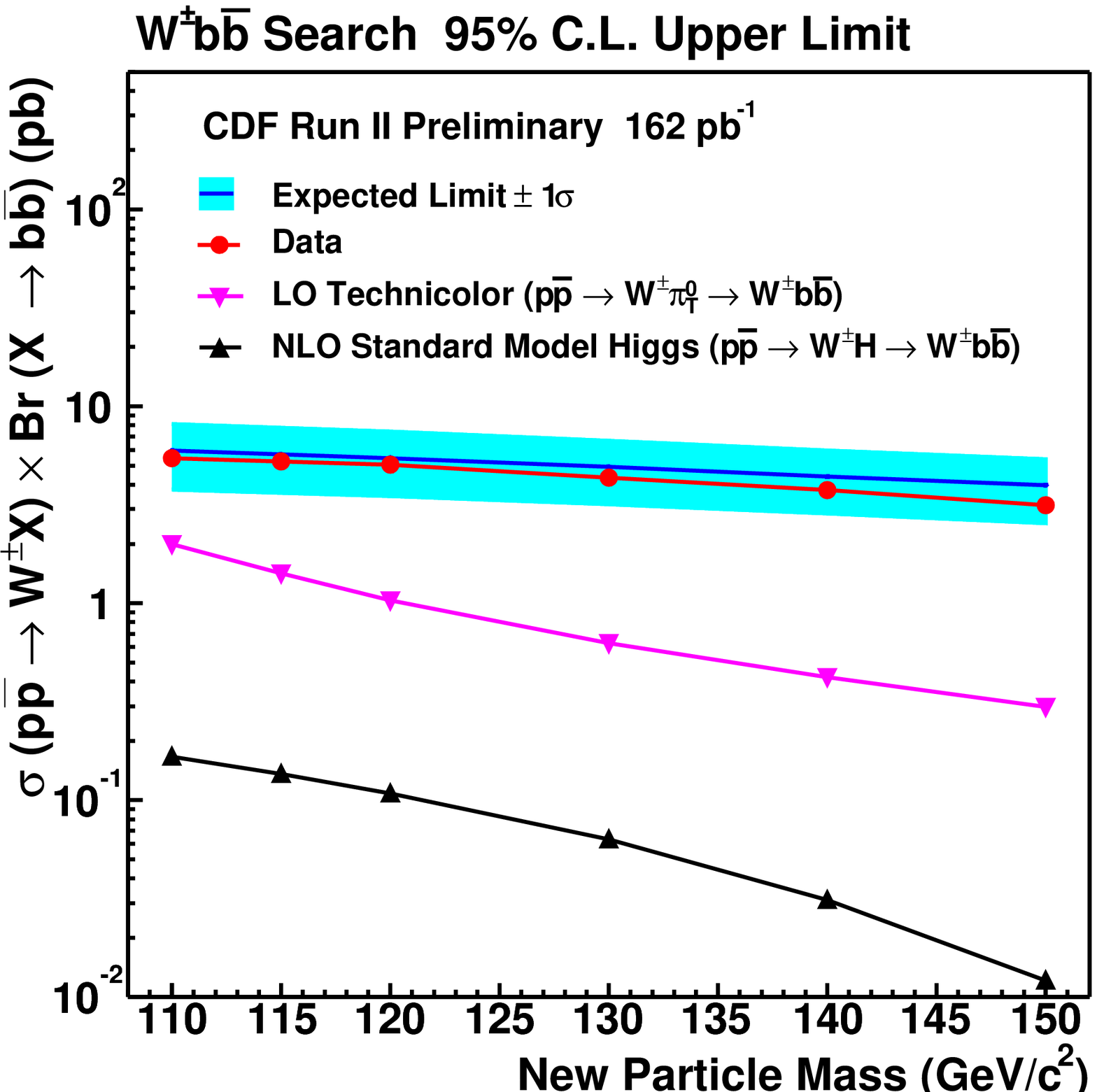,height=2.0in}
  \caption{
$WH$ results at CDF: a) Distribution of the dijet invariant mass in the $W+ $ 2 jets sample; b)
Same distribution, after requiring that at least one jet is $b$-tagged. The expectation 
obtained ($\times 10$)
from a 115 GeV Higgs is also shown. c) Limit on $WH~\times~BR(H \rar b \bar{b})$
cross section, as a function of $m_H$. 
     \label{fig:wh-cdf} }
\end{figure}

CDF has done a similar search  on 162 pb$^{-1}$ of data,
using a sample with one electron or muon ($p_T > 20$ GeV),
 2 jets ($p_T > 15$ GeV, $\eta<2$), 
\MET $>$ 20 GeV, but requiring only 
one jet to be $b$-tagged. The number of observed events goes from  $2072$ in the 
$W+2 $ jets sample (Fig.~\ref{fig:wh-cdf}a)
 to 62 events when requiring at least 1 $b$-tagged event  (Fig.~\ref{fig:wh-cdf}b).
The search for the Higgs is performed in a mass window on this distribution, 
normalizing the background to the data outside the window.
The resulting 95\% limit is  5-6 pb in the 105-135 GeV range.

\subsection{Search for  Higgs Production in 
Association with a $Z$ boson, in the $\nu \bar{\nu} b \bar{b}$ Final State}

\begin{figure}[bhp]
\psfig{figure=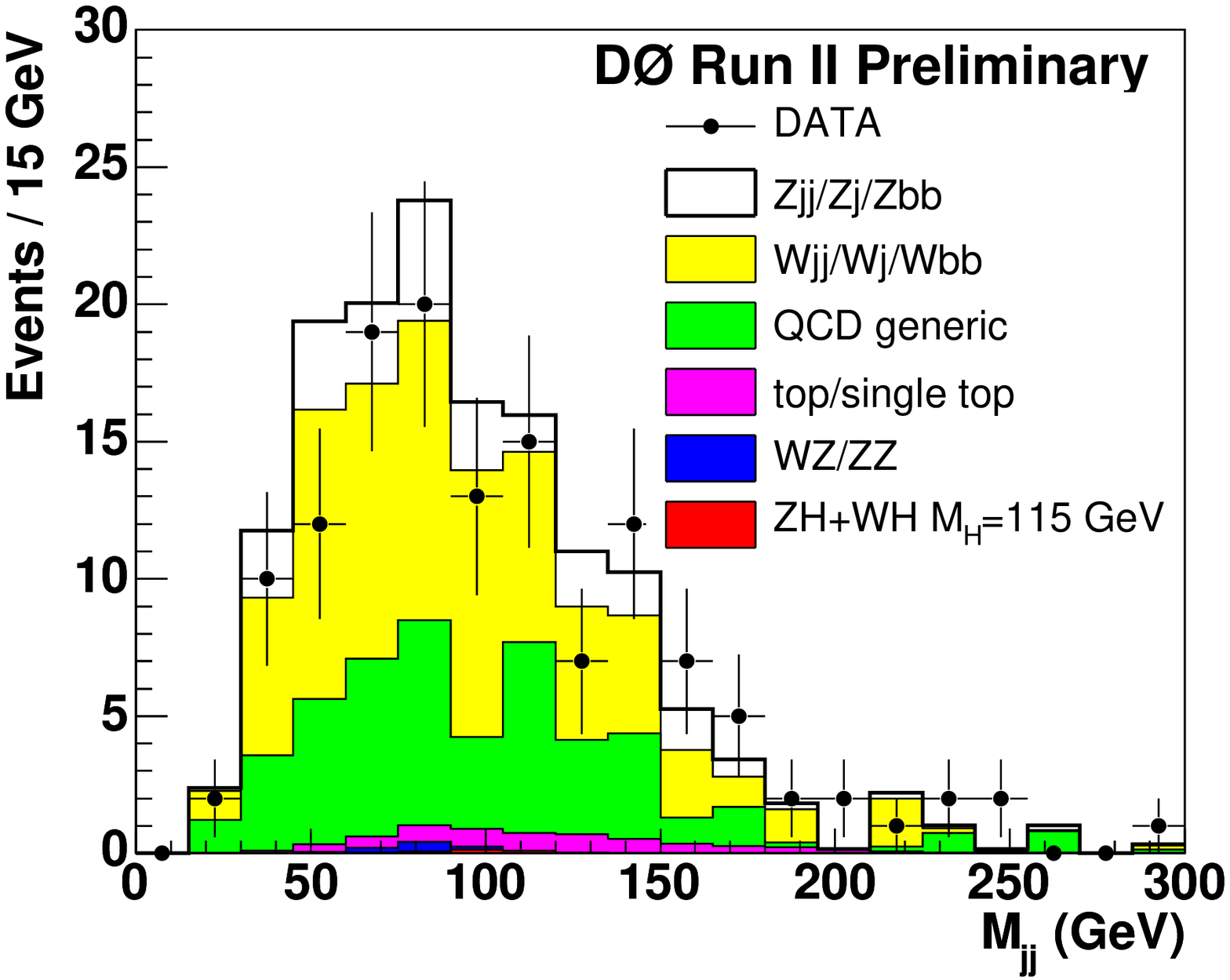,width=2.in,height=2.1in}
\psfig{figure=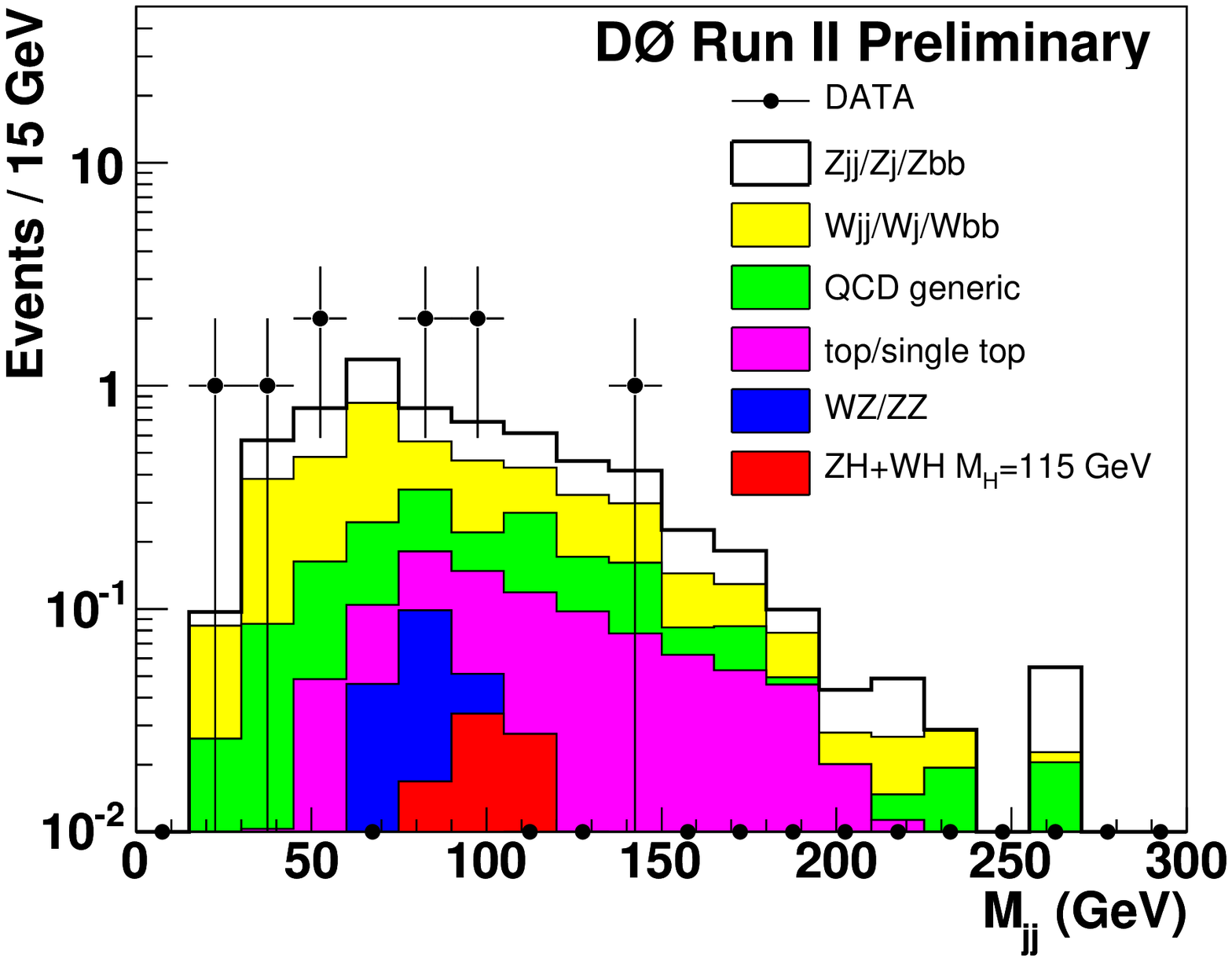,width=2.in,height=2.1in}
\psfig{figure=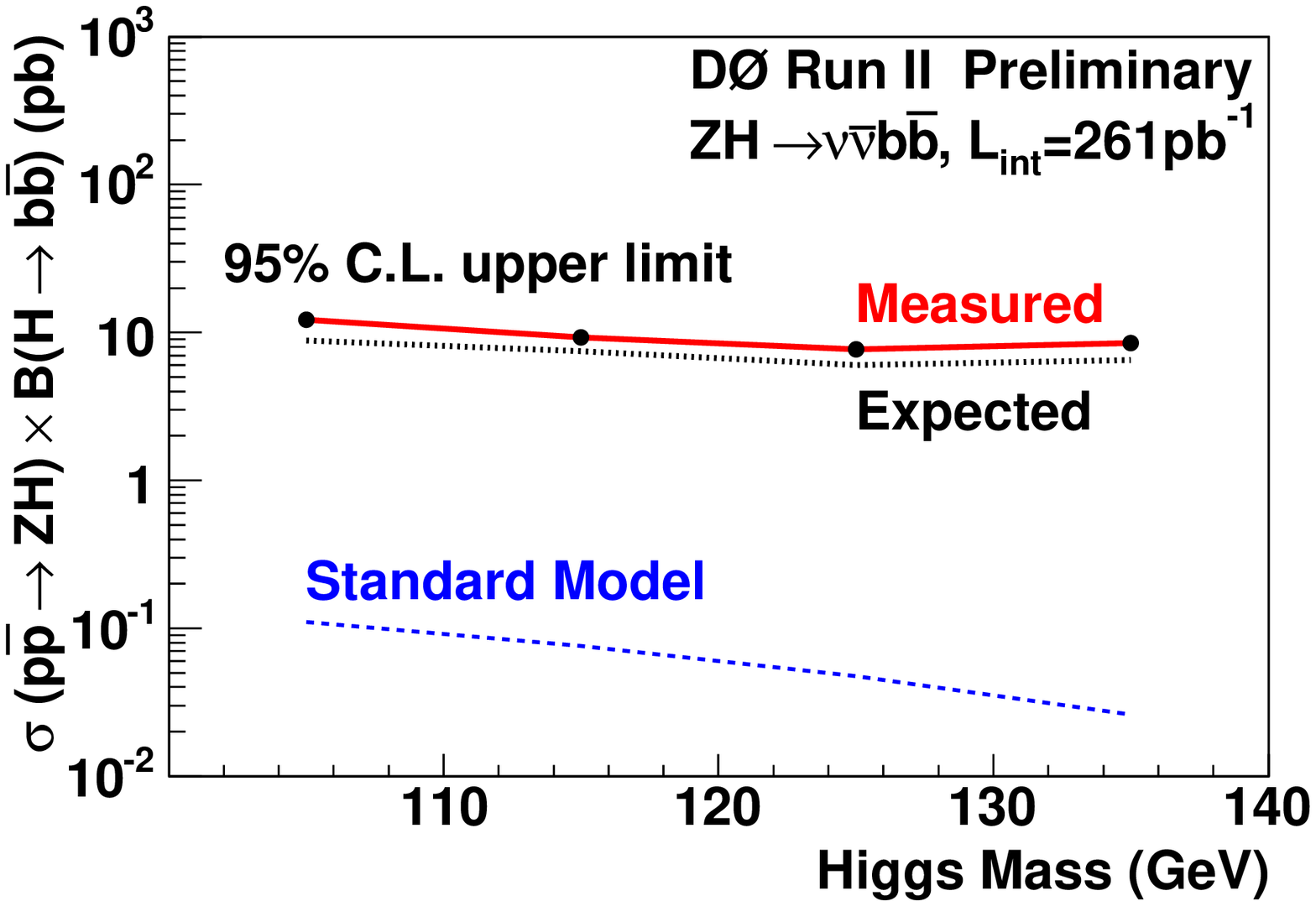,height=2.1in,width=2.4in}
  \caption{
ZH results at D\O .
a) Distribution of the dijet invariant mass in the $Z+ $ 2 jets sample after requiring at least one
$b$-tagged jet b)
Same distribution, after requiring that the 2 jets are $b$-tagged. The expectation obtained
from a 115 GeV Higgs is also shown. c) Limit on $ZH~\times~BR(H \rar b \bar{b})$
cross section, as a function of $m_H$. 
     \label{fig:zh-d0} }
\end{figure}

D\O\ has studied the  \MET + $b\bar{b}$ final state, in which 
$ZH (\rar \nu \bar{\nu} b \bar{b})$  production
could be observed~\footnote{This result became public after this conference (on 11/4/2005),
but is reported here for completeness.}.
The analysis is based on
261 pb$^{-1}$ of data, obtained using a dedicated trigger  selecting
events with acoplanar jets and \MET.
The final selection
which requires two $b$-tagged jets with 
$p_T >$ 20 GeV and $\eta < 2.5$, large transverse missing energy
(\MET $> 25$ GeV), no back-to-back topology, and no isolated tracks
allows to reject multijet background and "leptonic" $W+$ jets or $Z+$ jets events
in which the leptons escaped detection. The requirement $H_T < 200$ GeV, where $H_T$
is defined as the scalar sum of the \pt\ of the jets, allows to reject $t \bar{t}$ 
events. Further rejection of the multijet background is obtained by putting requirements
on the correlations between different ways of calculating \MET\ (using calorimeter cells,
tracks, or jets) which must be highly correlated in the case of the clean $\nu \bar{\nu}
b \bar{b}$ topology.
The measured dijet mass distribution is well described by the
expectation, as shown after single $b$-tagging  in 
Fig.~\ref{fig:zh-d0}a.  
The number of observed events is reduced from
 132 (2 jets including at least 1 $b$-tagged jet),
to 9 when requesting 2 $b$-tagged jets (Fig.~\ref{fig:zh-d0}b),
for respective SM expectations of 144.7 and 6.4 events. 
The dominant systematic uncertainties on the signal 
are due to the $b$-tagging efficiency ($\sim 22\%$)
and to the jet reconstruction/energy scale ($\sim 11\%$).
By restricting the selection to 
a $\pm$25 GeV window around the searched Higgs masses, 
D\O\ has established a 95\% C.L. limit on the $ZH~\times~BR(H \rar b \bar{b})$
production cross section, between 12.2 and 8.5 pb for $m_H$ between 105 and 135 GeV, 
as shown in Fig.~\ref{fig:zh-d0}c. The corresponding expected limits (8.8--6.5 pb)
are also shown.

This limit is, as expected, still significantly above the expected SM cross section.
For this channel and the current luminosity, 
the  sensitivity of this analysis due to the trigger/reconstruction
efficiencies compared to  the '03 prospective study, is approximately
a factor 3 lower.
As in the $WH$ case, work is in progress to approach the performance assumed
in the sensitivity study.

\subsection{Search for the SM Higgs Boson $H\rightarrow WW^{*} \rightarrow
l^{+}\nu l^{\prime -}\overline{\nu}$}

\begin{figure}[bhtp]
\psfig{figure=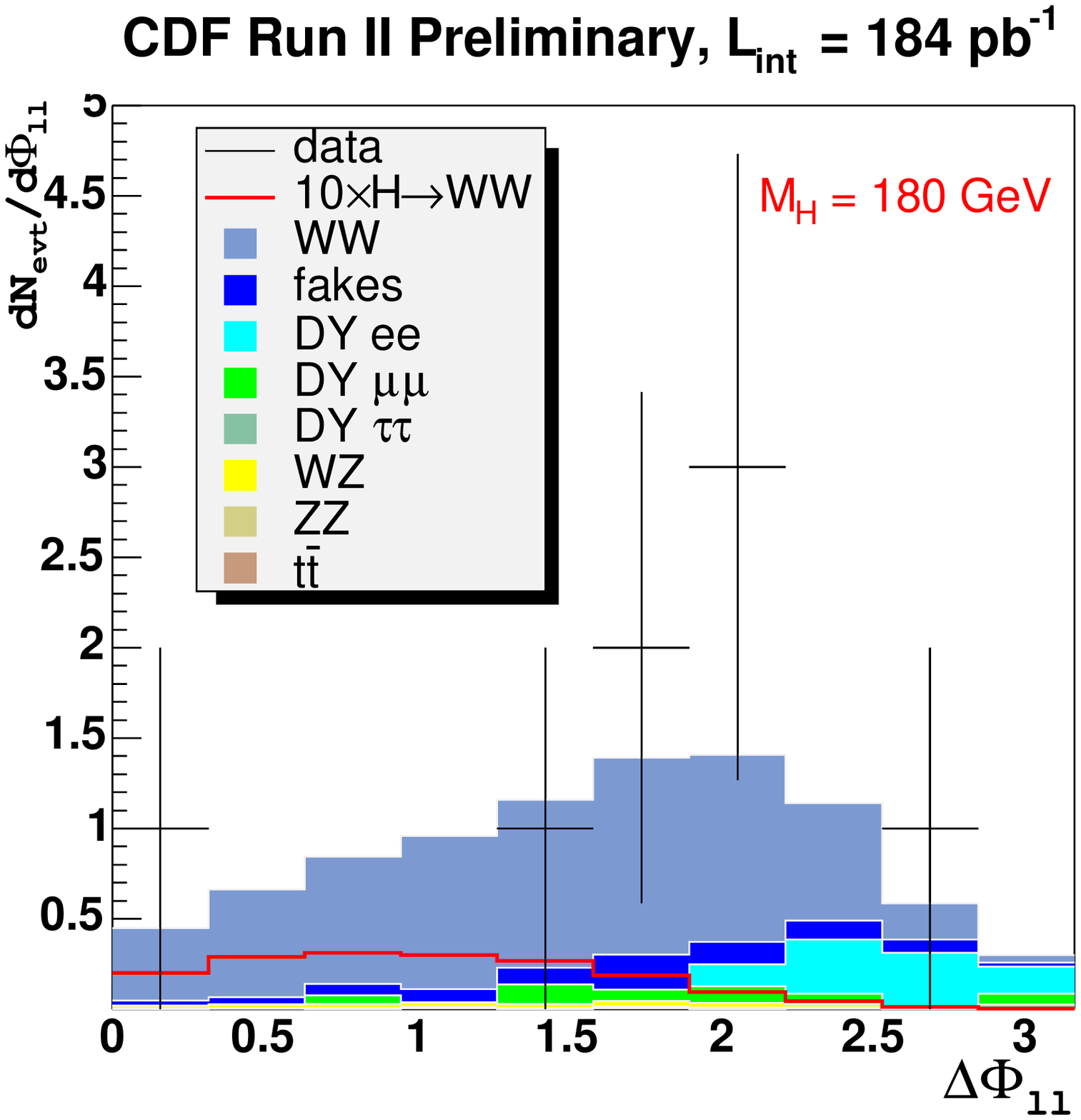,height=2.0in}
\psfig{figure=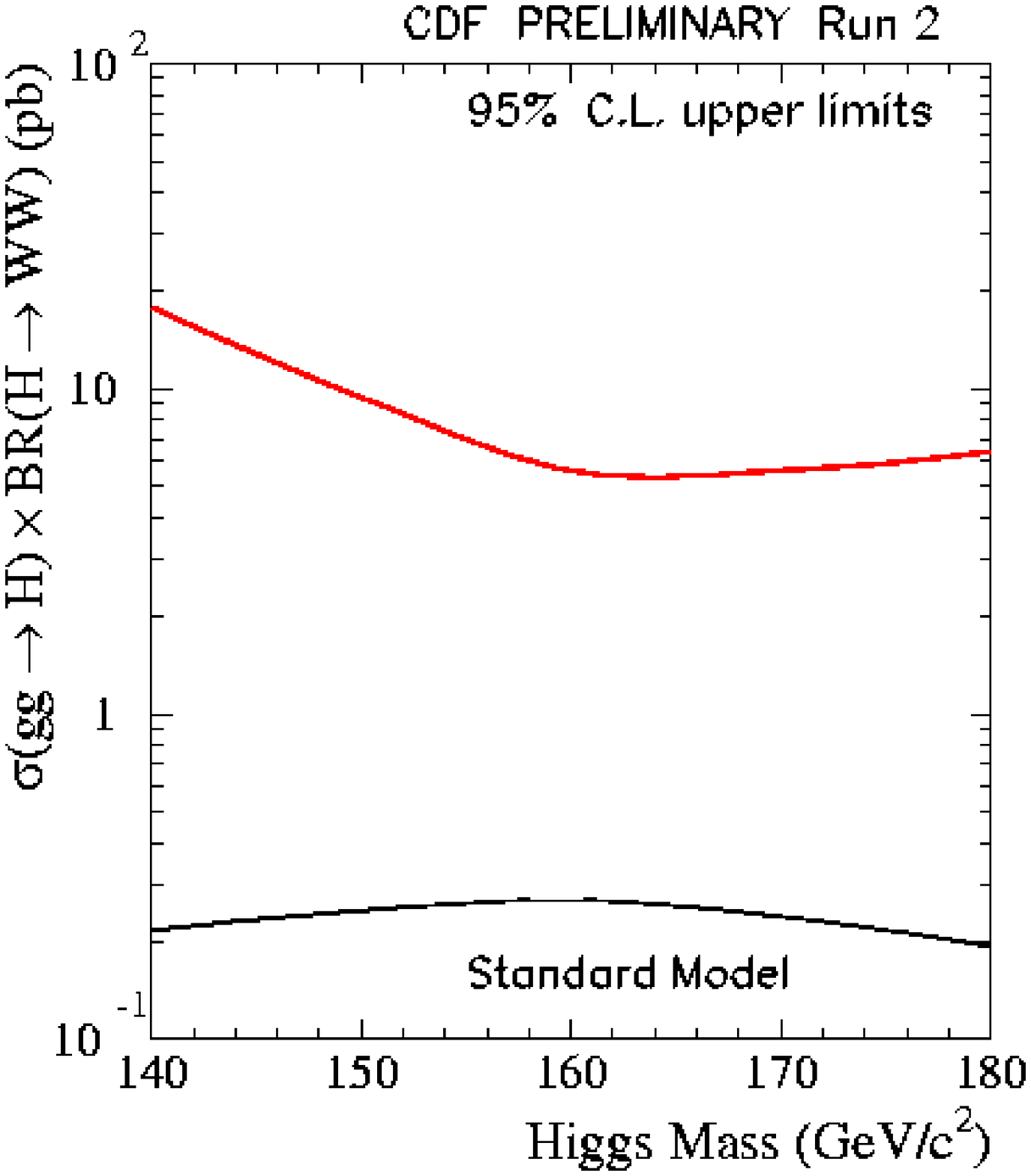,height=1.9in,width=2.0in}
\psfig{figure=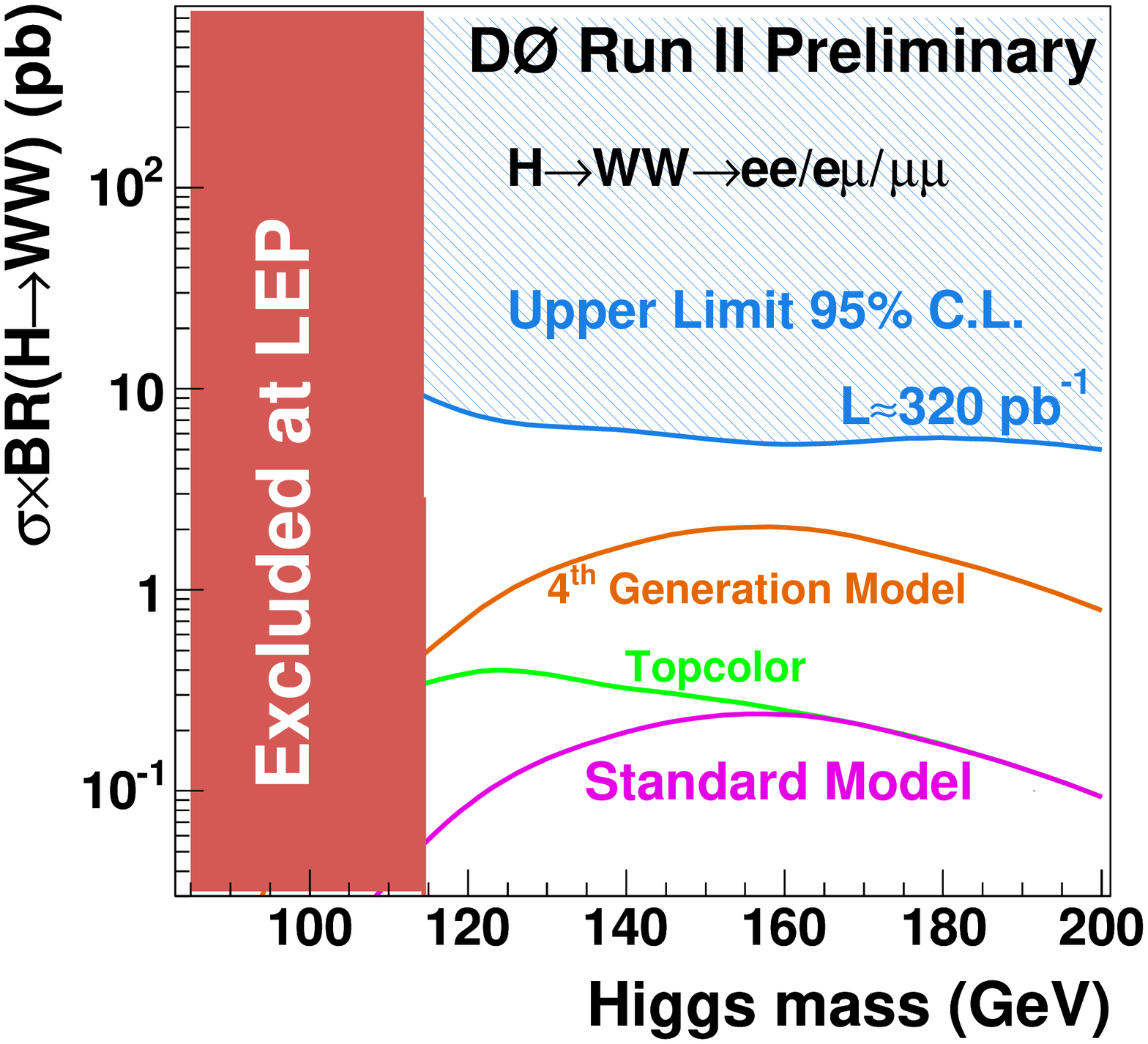,height=1.9in,width=2.2in}
  \caption{
$WW^*$  results at CDF and D\O . a) Distribution of $\Delta \varphi$ between the 2 leptons.
The $WW^*$ signal tend to be at lower $\Delta \varphi$ values. Limits on $WW^*$ production $\times$
$BR(H \rar WW^*)$ at CDF (b) and at D\O\ (c).
     \label{fig:ww-cdf-d0} }
\end{figure}

At higher Higgs masses, the search is best performed in the  $H\rightarrow WW^{*}$
channel. After measuring the $WW^{*}$ production cross section~\cite{d0-ww,cdf-ww},
which is well reproduced by Next-to-Leading Order calculations, 
CDF and D\O\ have searched for the SM Higgs decaying into $WW^{*}$  
decaying leptonically
($ee$, $e\mu$ and $\mu\mu$ channels).
The event selection requires 2 high \pt\ leptons of opposite charge, with an invariant
mass incompatible with the $Z$ mass and constrained
by the searched Higgs mass, and a significant \MET . Further requirements are based
on the sum of \MET\ 
 and  $p_T$ of the leptons, and on the $\varphi$ separation between leptons 
which are not back-to-back in Higgs decays due to spin correlations, as visible in
the CDF $\Delta \varphi$ distribution, Fig.~\ref{fig:ww-cdf-d0}a.

For CDF, on 184 pb$^{-1}$ of Data,
 8 events are  
observed, for an expectation of 9.1 events. The 95\% C.L. upper limit
on the cross-section times $BR(H \rar WW*)$
 is measured for masses between 140 and 180 GeV and
shown in Fig.~\ref{fig:ww-cdf-d0}b. For $m_H=160$ GeV, the limit is 
5.6~pb i.e. about an order of 
magnitude above the SM prediction.

For D\O\, on 320 pb$^{-1}$ of Data,
20 events are observed, 
for an SM expectation of 17.7 
events. In Fig.~\ref{fig:ww-cdf-d0}c is shown
the corresponding limit 
which  is 5.3 pb for $m_H$=160 GeV. Models assuming a 4$^{th}$ generation could be 
excluded in the near future, if $m_H$ is approximately 160 GeV.

\subsection{Search for the SM Higgs Boson in the $WH\rightarrow WWW^{*}$ channel}

\begin{figure}[htbp]
\psfig{figure=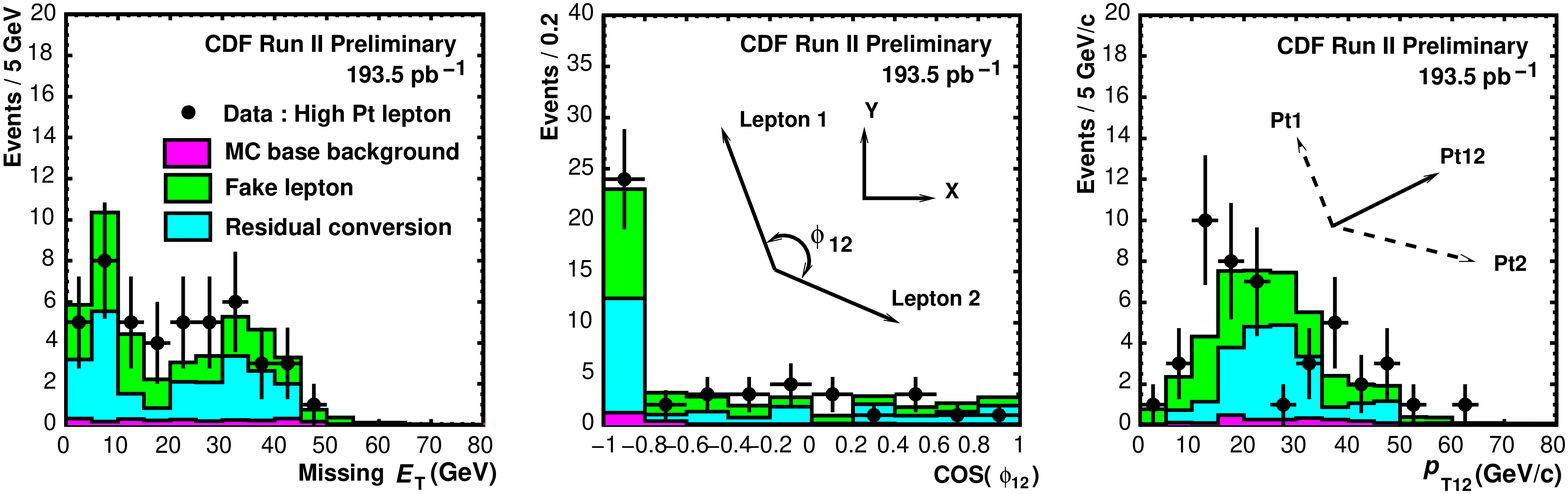,height=4.0in,bbllx=0,bblly=0,bburx=800,bbury=800,clip=}
\psfig{figure=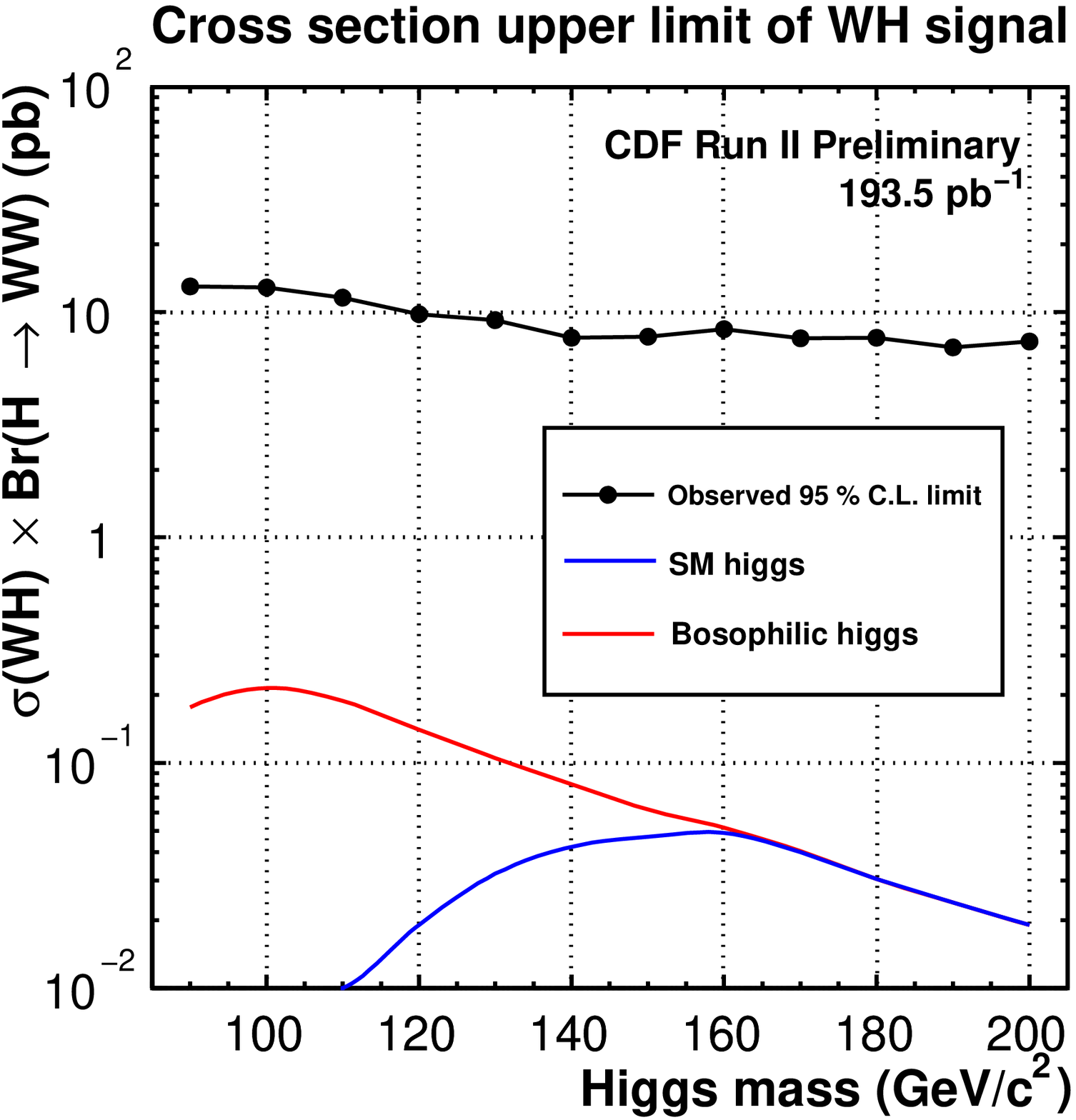,height=1.9in,width=2.2in}
  \caption{
$WWW^*$ results at CDF. Distributions of \MET\ (a) and $\cos {\varphi_{12}}$ (b) in the control sample (see
text); c)~95~\% CL limit obtained on $WWW^*$ production, 
compared to the SM expectation, and to the beyond the SM scenario of a bosophilic Higgs. 
     \label{www-cdf} }
\end{figure}

CDF has searched for $W$ Higgs production, in which $H \rar WW^*$,  using high-$p_T$
isolated like-sign dilepton events in  193.5 pb$^{-1}$ of data. The 
background components is studied in a like-sign sample selected
 by requiring the leading lepton $p_T >$  20 GeV and the second lepton
 $p_T >$
6 GeV. The entire sample is consistent with the
background expectation as shown in Fig.~\ref{www-cdf} for 2 typical distributions.
The signal region is determined in the plane
of the second lepton $p_T$ ($p_{T,2}$) versus the vector sum of $p_T$'s of the two
leptons($p_{T,12}$): $p_{T,2} >$ 16 (18) GeV and $p_{T,12} >$ 
35 GeV for
 Higgs masses $<$ 160 GeV ($>$ 160 GeV). No event is found, while the
total background is expected to be 0.95 $\pm$ 0.64  events. The expectation for a 110
GeV bosophilic  Higgs is about 0.06 events assuming the same
production cross section as the Standard Model Higgs, and for $m_H= $ 160 GeV, the 
SM Higgs expectation is about 0.03 events. For this mass, the 
cross section upper limits $\sigma(WH) \times
BR(H\rar WW^*)$ is $\sim 8$ pb, as shown in Fig.~\ref{www-cdf}c.

\section{Searches for Supersymmetric Higgs Bosons}
Searches  for non-SM Higgs bosons are also performed at the Tevatron.
CDF and D\O\ have started searching for exotic Higgs bosons 
including SUSY Higgs bosons and doubly charged 
Higgs bosons. In the case of SUSY, the efforts have concentrated so 
far on the minimal supersymmetric standard model 
(MSSM) Higgs bosons.


In two Higgs-doublet models of electroweak symmetry breaking, such
as in the MSSM~\cite{susy}, there are five physical Higgs bosons:
two neutral $CP$-even scalars, $h$ and
$H$; a neutral $CP$-odd state, $A$; and two charged states, $H^\pm$. The ratio of the vacuum
expectation values of the two Higgs fields is defined as \tanb =
$v_2/v_1$, where $v_2$ and $v_1$ refer to the Higgs fields that couple to
the up-type and down-type fermions, respectively.
At tree level, the coupling of the $A$ boson to down-type quarks,
such as the $b$ quark,
is enhanced by a factor of \tanb\ relative to the standard model (SM),
and the production
cross section is therefore enhanced by $\tan^{2}\beta$~\cite{MSSM_Higgs}.

LEP experiments have excluded at 95\% C.L. a light Higgs boson with mass
$m_h$ $\le$ 92.9~GeV~\cite{leplimit}, independent of \tanb,
and at higher masses for \tanb\ below 10--20.
The Tevatron is currently sensitive to \tanb\ values between $\sim$ 50 and $\sim$ 100.
In this region of \tanb, the $A$ boson is nearly degenerate in
mass with either the \hboson\ or the \Hboson\ boson, and their
widths are small compared to the dijet mass resolution.
Since we cannot distinguish between the \hboson/\Hboson\
and the \Aboson, the total cross section for signal is
assumed to be twice that of the $A$ boson.

The major decay modes of $h$ 
are a $b$ quark pair production($\sim 90\%$) and $\tau$ lepton pair production($\sim 8\%$).
The $p\bar{p} \rightarrow hb\bar{b}$ with $h\rightarrow b\bar{b}$ 
or the gluon-gluon higgs production with 
$h \rightarrow \tau\tau$ decay are the most promising 
channels at the Tevatron. 

\subsection{Search for MSSM Higgs Bosons in the $hb\bar{b}$ Channel}

D\O\ has searched for the MSSM Higgs in the $hb\bar{b}$ channel using
 260~$\mbox{pb}^{-1}$ of data, with a
dedicated trigger based on high ($\geq 3$) jet multiplicity, 
designed for maximizing signal acceptance 
while providing acceptable trigger rate.
Events with up to five jets were initially selected, while the signal was
searched as a mass resonance in the subsample having at least 
3 $b-$tagged jets. 
Jets containing \btag\ quarks were identified using a secondary
vertex (SV) tagging algorithm. 
The \btag\ tagging efficiency is $\approx$~55\% for central
\btag-jets of \pt$>$35~GeV with a light quark (or gluon) mistag rate
of about 1\%.

The  $b\bar{b} h$ signal events, with $h\rightarrow b\bar{b}$, were 
generated for Higgs boson masses from 90 to 150~GeV using 
{\footnotesize PYTHIA} event generator~\cite{pythia}.
The \pt\ and rapidity spectra of the Higgs bosons from 
{\footnotesize PYTHIA} were normalized to those from NLO calculation~\cite{5fns}.
The multijet production is the largest
background and is determined from data 
but is also verified using the {\sc alpgen}~\cite{alpgen} event generator.
All other backgrounds are expected
to be small and are simulated with {\sc pythia}.

\begin{figure}[!htpb]
\begin{center}
\psfig{figure=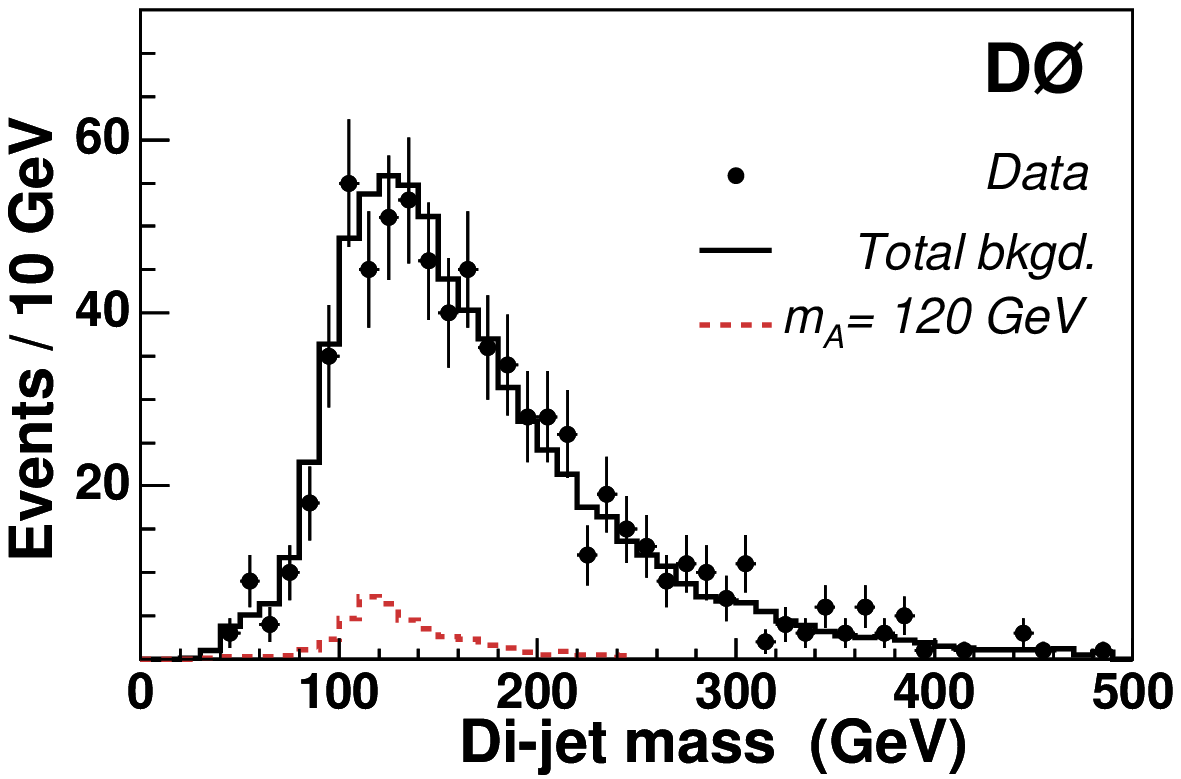,height=2.in,width=2.in}
\psfig{figure=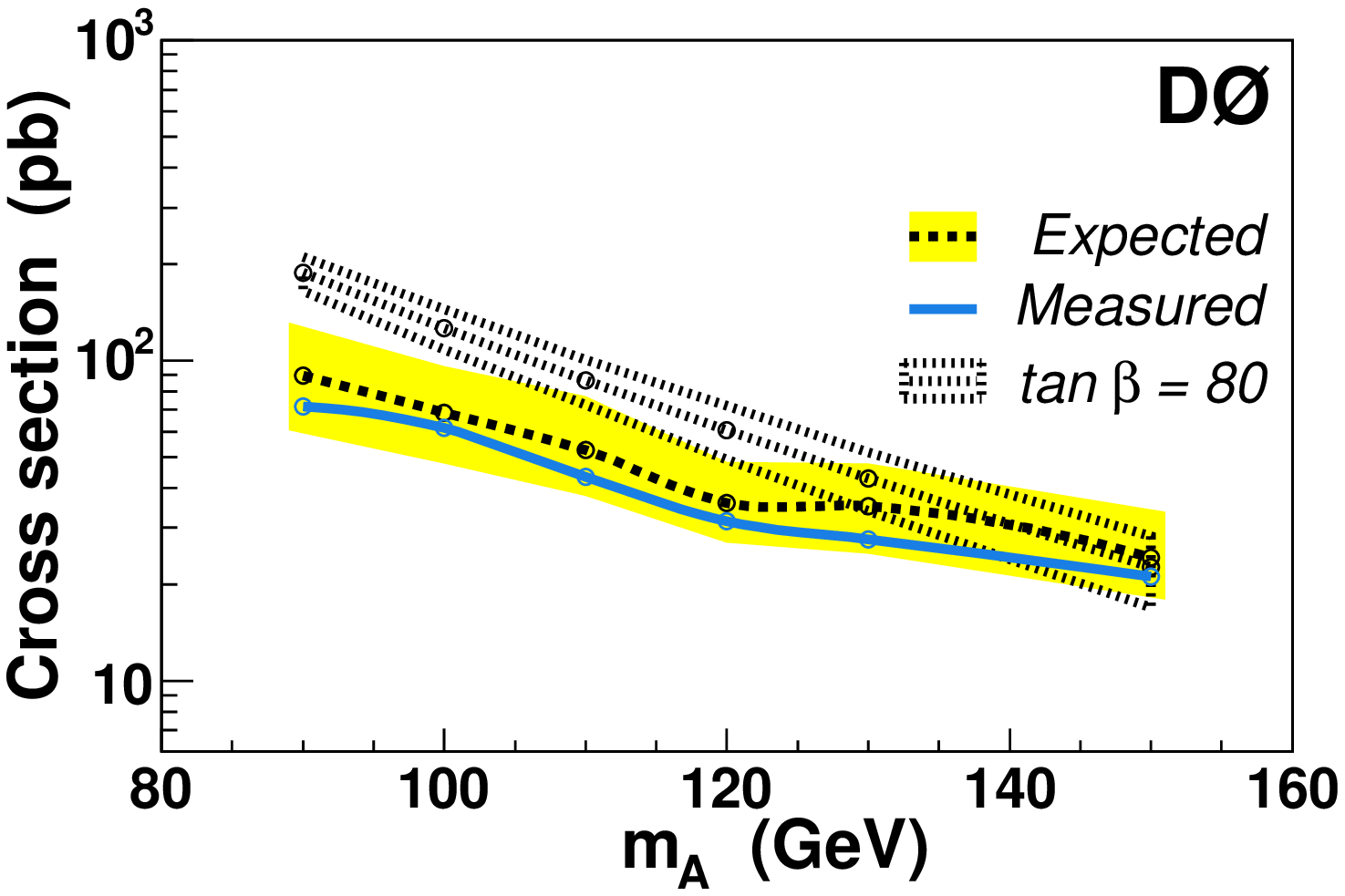,height=2.in,width=2.in}
\psfig{figure=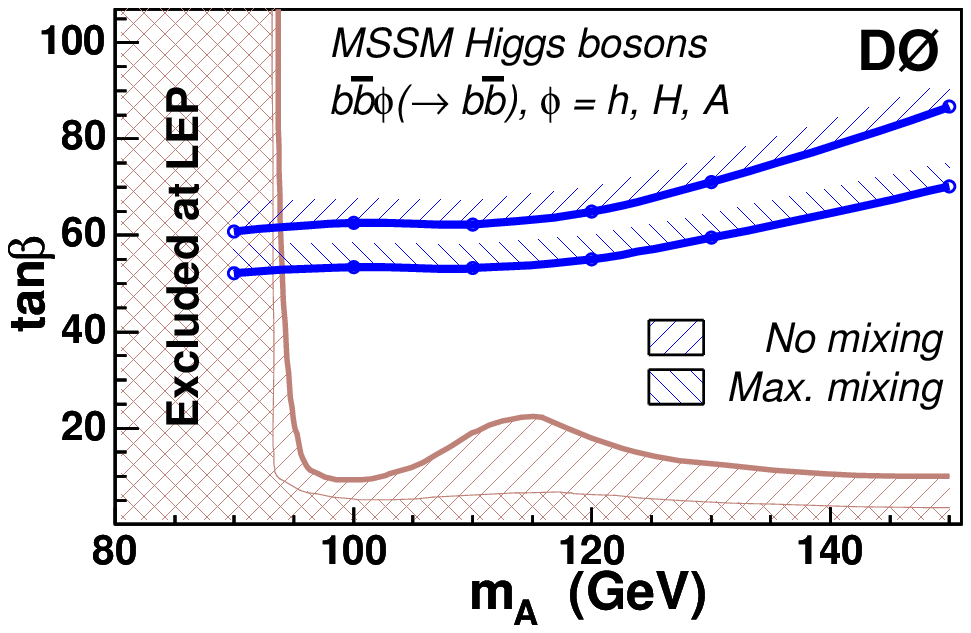,height=2.in,width=2.2in
}
\end{center}
\caption{$hb \bar{b}$ results at D\O .
a) Dijet mass distribution obtained from the 2 leading $b$-tagged jets 
in multijet events with at least 3 $b$-tagged events. b)
Upper limits on the signal cross section as a function of Higgs boson mass.
c) 95\% C.L. upper limit 
on $\tan{\beta}$ as a function of Higgs boson mass~(right).} \label{fig:mh_limit}
\end{figure}

There are two main categories of multijet background. One contains
genuine heavy-flavor jets, while the other consists of light-quark
or gluon jets that are mistakenly tagged as \btag-quark jets, or
correspond to gluons that "split" into nearly collinear \bbbar\
pairs. Using the selected data sample, before the application of
\btag-tagging requirements, the probability to \btag-tag a jet (``mistag'' function) was
measured as a function of its \pt\ and $|\eta |$. 
The mistag function was corrected by subtracting heavy-flavor contributions.
and used to estimate
the instrumental background, by applying them to every jet
reconstructed in the full data sample.
The triple $b$-tagged data were compared with this background expectation,
as shown in Fig.~\ref{fig:mh_limit}a.
No evidence for signal was found in mass windows centered around the searched
Higgs masses. The expected dijet mass distribution originating from a
120 GeV MSSM Higgs boson is also shown in the
figure.

Figure~\ref{fig:mh_limit}b shows the expected MSSM Higgs boson
production cross section as a function of $m_A$ for \tanb = 80. 
The MSSM cross
section shown in Figure~\ref{fig:mh_limit} corresponds to no mixing in
the scalar top quark sector~\cite{tev}, or $X_{t}$ = 0, where
${X_{t} = A_{t} - \mu\cot\beta}$, $A_{t}$ is the tri-linear
coupling, and the Higgsino mass parameter $\mu = -0.2$~TeV. The results are 
also 
interpreted in the ``maximal mixing'' scenario with $X_{t}
= \sqrt6 \times M_{{SUSY}}$, where $M_{{SUSY}}$ is the
mass scale of supersymmetric particles, taken to be 1~TeV.
Results for both scenarios of the MSSM are shown in
Figure~\ref{fig:mh_limit}c as limits in the \tanb\ versus $m_A$ plane.
The present D\O\ analysis excludes a
significant portion of the parameter space, down to \tanb\ = 50,
depending on $m_A$ and the MSSM scenario assumed.

\subsection{Search for MSSM Higgs Bosons in the $gg \rar hX \rightarrow \tau \tau X$ Channel}

Using  an integrated luminosity of about 200 $\mbox{pb}^{-1}$, 
CDF has searched for a MSSM Higgs  
in the $gg \rar hX \rightarrow \tau \tau X$ channel
The signal consists of a tau pair in which one of the taus decays to hadrons and a 
neutrino ($\tau_h$) while the other decays to an electron ($\tau_e$) or a muon ($\tau_\mu$)
and  two neutrinos.
The data were collected with a set of dedicated $\tau$-triggers which 
select a lepton ($e$ or $\mu$) candidate with $p_T>$ 8~GeV and an isolated track from $\tau_h$ decay.
The $\tau$ lepton decaying into hadrons, was identified/reconstructed by 
the charged tracks and the neutral pions lying inside 
a narrow cone. The invariant mass  and 
the track multiplicity of the $\tau_h$ were required to be less than 1.8~GeV and equal to 
1 or 3, respectively.
Further selections were done on the event topology using 
missing transverse energy (\MET) and the $p_T$ of the taus. 
An example is the scalar sum $\hat{H_T} = |p_T^{vis}(\tau_1)|+|p_T^{vis}(\tau_2)|+$\MET, 
where $p_T^{vis}(\tau)$ is the visible transverse momentum of the tau decay products not including 
the neutrinos.
Requiring $\hat{H_T}>$50~GeV leads to significant background reduction with small signal loss.

\begin{figure}[!htpb]
\begin{center}
\psfig{figure=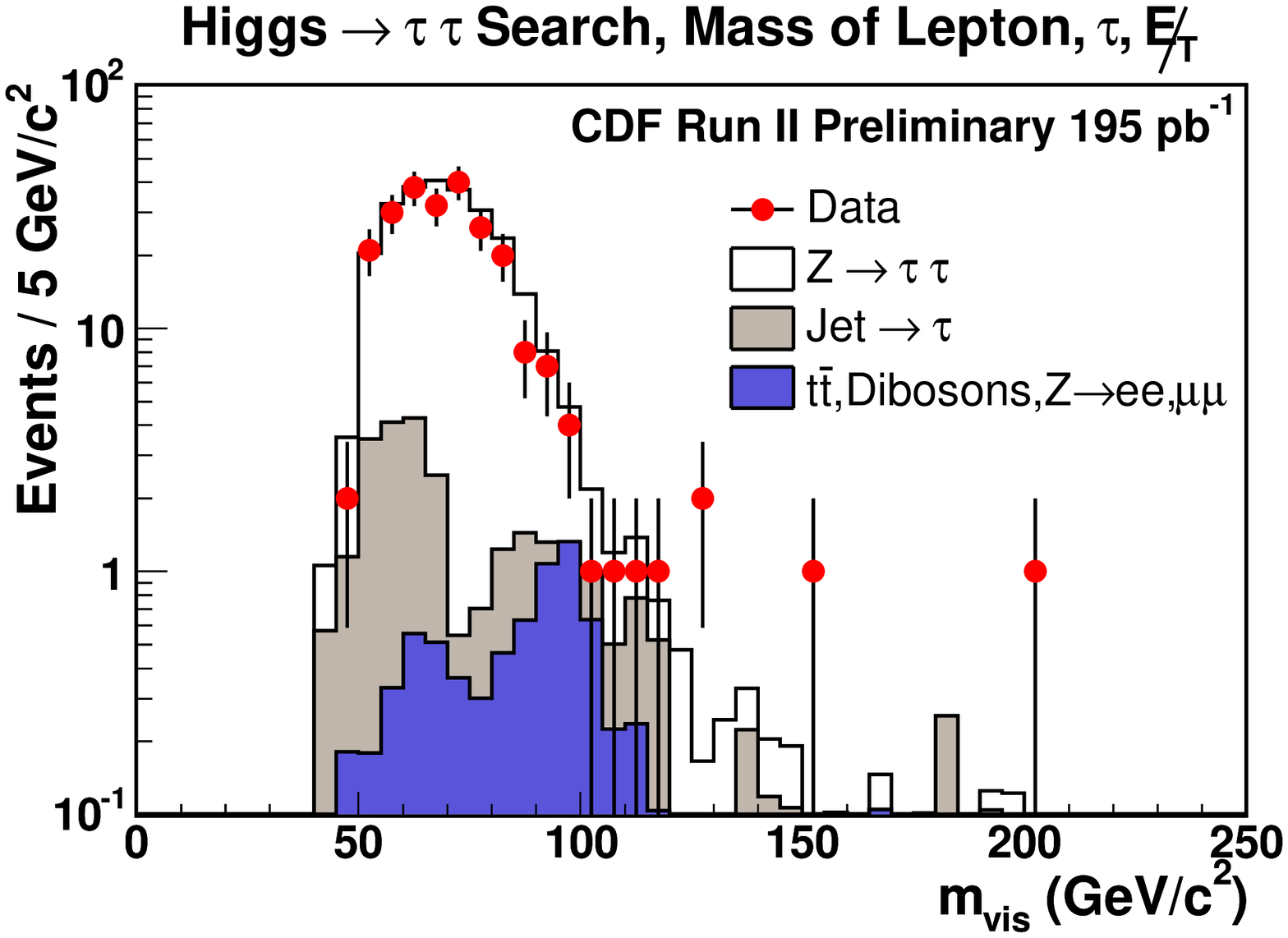,height=2.in,width=2.in}
\psfig{figure=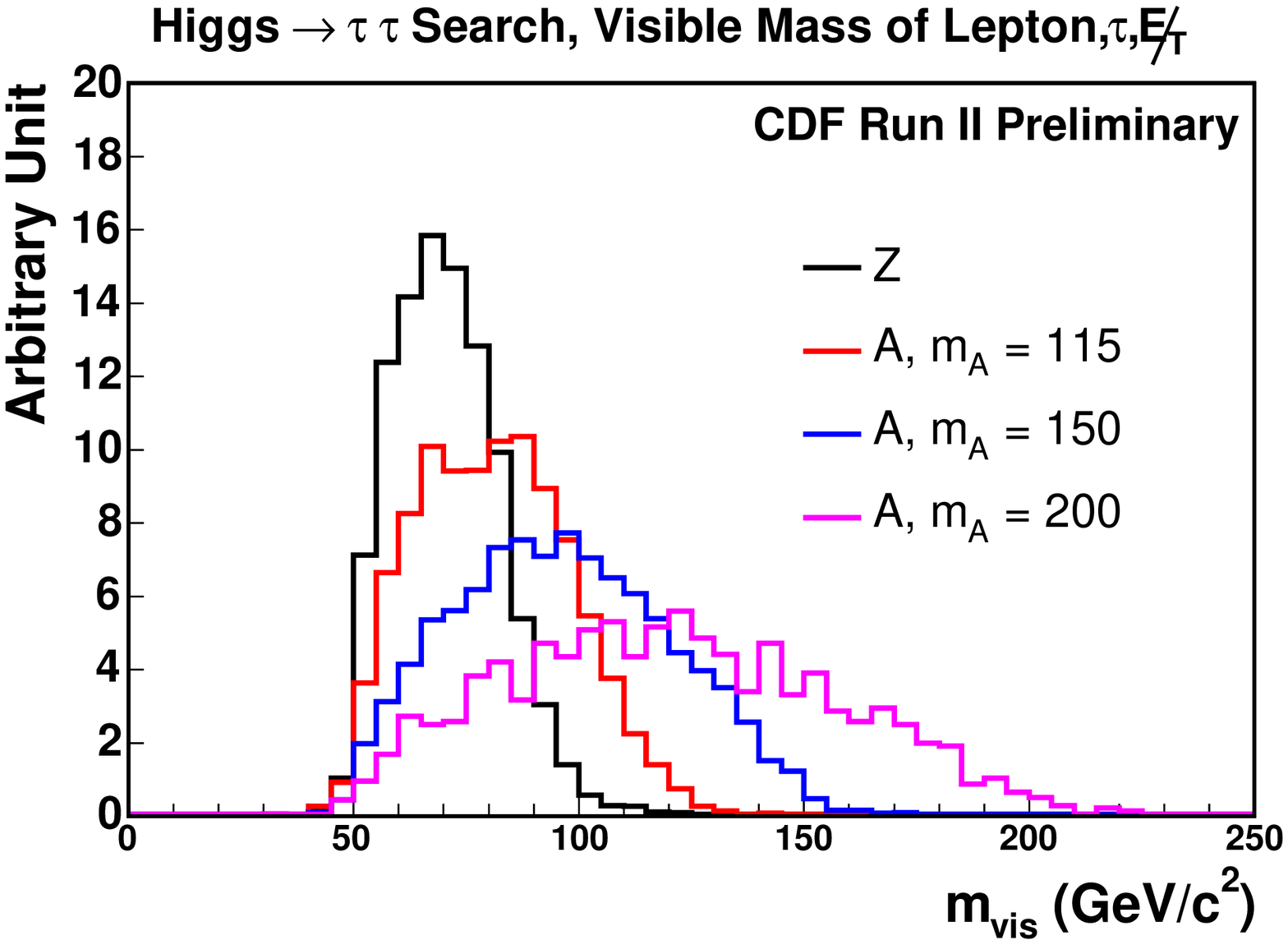,height=2.in,width=2.in}
\psfig{figure=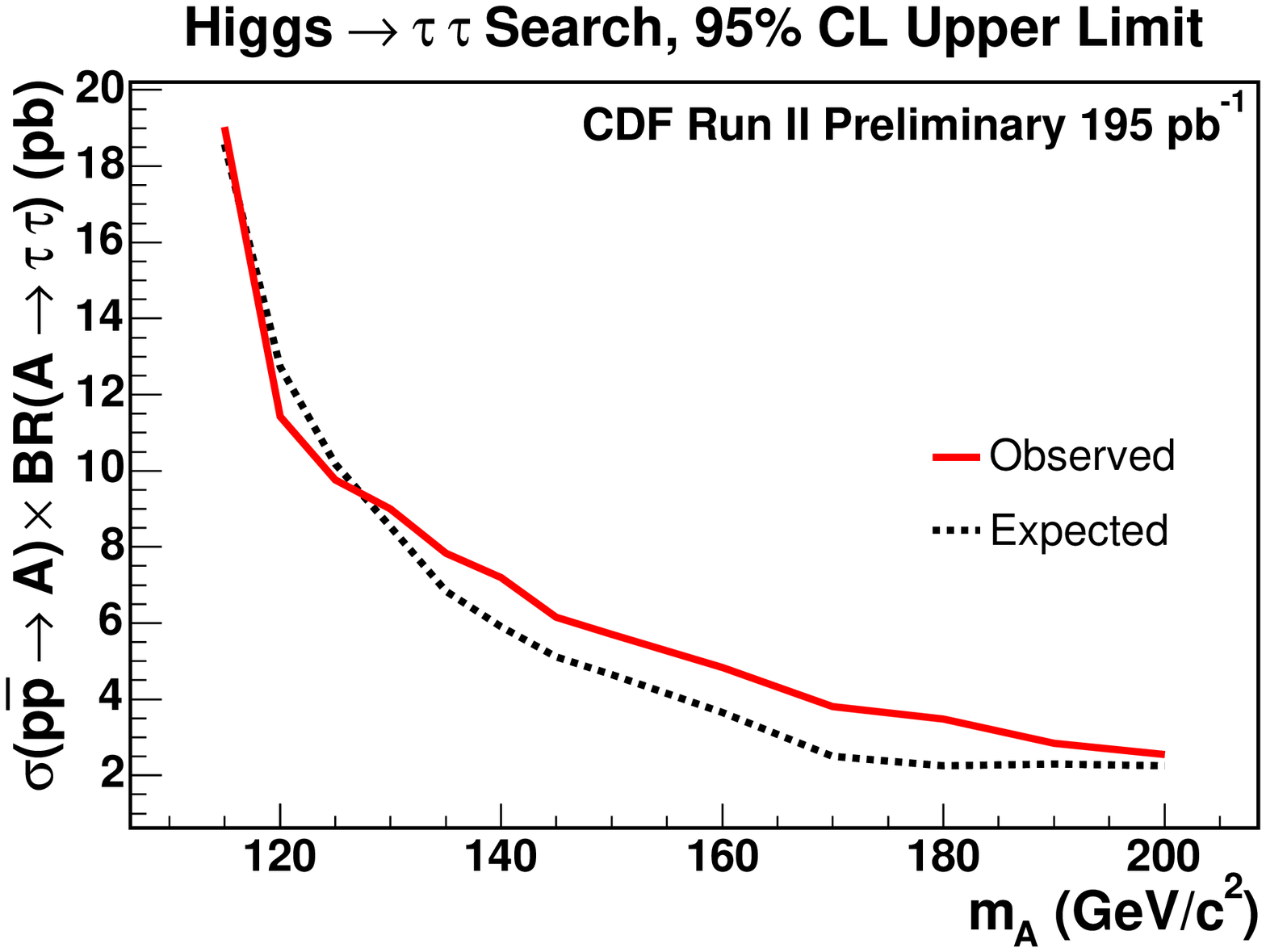,height=2.in,width=2.in}
\end{center}
\caption{$A  \rar \tau \tau $ results at CDF.
a) Observed visible invariant mass $m_{vis}$($l$,$\tau_h^{vis}$, \MET) distribution compared to
the sum of various backgrounds; b) comparison of $m_{vis}$($l$,$\tau_h^{vis}$) 
reconstructed with
$\tau$'s originating from $Z$ or $A$ decay; c) 95\% CL upper limit 
on
$\sigma(p\bar{p} \rightarrow h) \times B(A \rightarrow \tau\tau)$
as a function of $m_H$.} \label{fig:mass_fit}
\end{figure}

The signal processes $gg\rightarrow h$ and $b\bar{b}\rightarrow h$ were simulated using
{\footnotesize PYTHIA}. Higgs masses between 115~GeV and 
200~GeV were generated for \tanb $=$30. The backgrounds from $Z\rightarrow l^{+}l^{-}$, 
($l$=$e$,$\mu$,or $\tau$), di-boson production, and $t\bar{t}$ production 
were estimated using Monte Carlo samples. The backgrounds from jet to $\tau$ misidentification 
were estimated using a fake rate function obtained from independent jet samples, as a function 
of the jet energy, pseudorapidity, and track multiplicity. 
The biggest source of background is $Z\rightarrow \tau\tau$ events,
which can only be distinguished from the Higgs signal by the di-$\tau$  mass.
A mass-like discriminating variable, $m_{vis}$($l$,$\tau_h^{vis}$, \MET),
was constructed using the four-momentum of the lepton ($e$ or $\mu$), the 
four-momentum of the visible decay products of the $\tau_h$($\tau_h^{vis}$)
and \MET\ (also treated as a four vector). After combining the $\tau_e\tau_h$ 
and $\tau_h\tau_\mu$ channels, 
a binned likelihood fit of the $m_{vis}$($l$,$\tau_h^{vis}$, \MET)
distribution was done 
to search for the Higgs signal. 
In  Fig.~\ref{fig:mass_fit} is shown the $m_{vis}$($l$,$\tau_h^{vis}$, \MET) distribution 
for data and various backgrounds. 
The limits on Higgs production cross section times branching ratio 
at 95 \% C.L. are presented in Fig.~\ref{fig:mass_fit}. 

\section{Searches for Doubly Charged Higgs Bosons}

\begin{figure}[!htpb]
\begin{center}
\psfig{figure=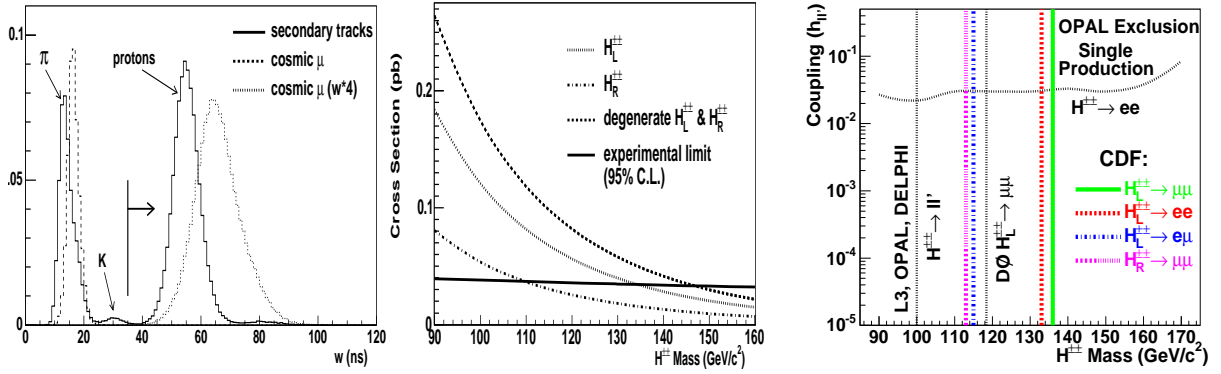,width=7.3in,bbllx=100,bblly=560,bburx=600,bbury=720,clip=}
\end{center}
\caption{$H^{\pm\pm}$ results at CDF and D\O .
a) Distribution of the CDF tracking detector dE/dx variable w for positively-charged secondary 
tracks with \pt\ of 300 -350 MeV/c (solid), for high-pT cosmic ray muons (dashed), 
and the expectation for H±± tracks (dotted); 
b)
comparison of the CDF experimental cross section upper limit with theoretical 
NLO cross section 
for pair production of long-lived $H^{\pm\pm}$ bosons, with 
 left-handed
($H^{\pm\pm}_L$) and right-handed ($H^{\pm\pm}_R$) couplings, 
and summed for the degenerate mass case;
c)
comparison of the CDF and D\O\ experimental cross section upper limit with theoretical 
NLO cross section 
for pair production of short-lived $H^{\pm\pm}$.}
\label{fig:dhiggs_limit}
\end{figure}

Several models like the left-right symmetric model~\cite{left-right}
require a Higgs triplet leading to an observable doubly-charged 
Higgs boson~($H^{\pm\pm}$), 
which could be light in the minimal supersymmetric left-right model.
The dominant production mode at the Tevatron is expected to be 
$p\bar{p} \rightarrow \gamma^{*}/Z + X \rightarrow H^{++} H^{--}$.
The partial width in the leptonic decay mode is proportional to the coupling to the 
lepton and to the Higgs mass.
D\O\ and CDF have already published mass limits from direct searches in the 
di-lepton decay channels for 
the short lived $H^{\pm\pm}$~\cite{dhiggsSL_D0}$^,\ $\cite{dhiggsSL_CDF}, as shown
in Fig.~\ref{fig:dhiggs_limit}a. The LEP
 experiments 
exclude $m_{H^{\pm\pm}} <\sim $ 99.5~GeV
at 95\% C.L. for $H^{\pm\pm}$ bosons with couplings to left- or right-handed leptons.

If the $H^{\pm\pm}$ boson is long-lived~($c\tau>$3~m), 
it can decay outside the detector. 
CDF has performed such a search  on 292 $\mbox{pb}^{-1}$ of data~\cite{dhiggsLL_CDF}, using
an inclusive muon trigger which requires a track with  $p_T>$ 18~GeV, and a match in the muon chamber.
The event selection requires two tracks, each with $p_T>$20~GeV, and at least one of them must have 
a matching muon. Since the charge
collected by the drift chamber is proportional to the ionization deposited by 
the particle per unit length($dE/dx$), an $H^{\pm\pm}$ boson would deposit a $dE/dx$ 
four times larger than a single charge tracks like an isolated $\mu$. 
The $H^{\pm\pm}$ boson charge deposit is modelled by quadrupling the $dE/dx$ of cosmic ray muons, while
the $dE/dx$ for low momentum protons are used as a controlled sample, see 
 Fig.~\ref{fig:dhiggs_limit}b.

Since no events remained after selecting large $dE/dx$, 95\% C.L. upper limits on 
the $H^{\pm}$ pair production cross section were set, as shown in 
Fig.~\ref{fig:dhiggs_limit}c.
The theoretical cross sections are computed separately for $H^{\pm\pm}$ bosons that couple 
to left- and right-handed particles ($H^{\pm\pm}_L$ and $H^{\pm\pm}_R$). 
Long-lived $H^{\pm\pm}_L$ and $H^{\pm\pm}_R$ bosons were excluded below 
a mass of 133~GeV and 109~GeV, respectively. 
When the two states are degenerate in mass, the exclusion limit becomes
$H^{\pm\pm}<$146~GeV.

\section{Conclusion}
We reported on searches for SM and beyond SM Higgs bosons. No evidence for Higgs signal 
has been found yet,
but the evolution of the Tevatron luminosity indicates that a light SM Higgs is within reach
for CDF and D\O\, by the years 2008-2009. To achieve this difficult goal, all channels will
have to be exploited and combined, 
and the first results obtained by the two collaborations allow to be
optimistic. A significant improvement of
sensitivity in the BSM Higgs searches is also taking place, 
as the first Tevatron Run II results are demonstrating.

\section*{Acknowledgments}
I would like to thank the organizers for an interesting conference and the European Community
for a Grant allowing European researchers working abroad to participate at this conference.
Many thanks also to my CDF and D\O\ colleagues working on this exciting topic who produced the results
presented in this summary. I would like to thank in particular St\'ephanie Beauceron, Suyong Choi, 
Beate Heinemann, Avto Kharchilava and Makoto Tomoto.

\end{document}